\documentclass[british,longbibliography,nofootinbib]{revtex4-2}
\usepackage[T1]{fontenc}
\usepackage[latin9]{inputenc}
\usepackage{color}
\usepackage{babel}
\usepackage{amsmath}
\usepackage{amssymb}
\usepackage{graphicx}
\usepackage[pdfusetitle,
 bookmarks=true,bookmarksnumbered=false,bookmarksopen=false,
 breaklinks=false,pdfborder={0 0 1},backref=false,colorlinks=true]
 {hyperref}
\hypersetup{
 pdfborderstyle=,citecolor={blue}}

\makeatletter
\usepackage{babel}

\makeatother

\begin{document}
\title{All-optical coherent control of chiral electronic transitions for
highly enantioselective photochemistry}
\author{Andrés Ordóñez$^{1,2,*,\dagger}$, Patricia Vindel-Zandbergen$^{3}$ and
David Ayuso$^{1,2,4,\ddagger}$}
\affiliation{$^{1}$Department of Physics, Imperial College London, SW7 2AZ London,
UK~\\
$^{2}$Department of Chemistry, Queen Mary University of London, E1
4NS London, UK~\\
$^{3}$Department of Chemistry, New York University, New York 10003,
USA ~\\
$^{4}$Max-Born-Institut, Max-Born-Str. 2A, 12489 Berlin, Germany~\\
$^{*}$andres.ordonez@fu-berlin.de
}
\begin{abstract}
Enantioselective photochemistry provides access to unique molecular
structures and functions, with deep implications for fundamental science
and industrial applications. Current methods for highly enantioselective
photochemistry critically rely on chiral sensitisers, as circularly
polarised light on its own yields vanishingly weak enantioselectivity.
 Here, we introduce a quantum control strategy to drive highly enantioselective
electronic excitations in randomly oriented samples using a pulsed
($\sim$22 fs) IR laser and two of its harmonics, in the absence of
intermediate resonances. Our approach addresses electronic transitions,
does not require chiral sensitisers, or cold molecules, or long electronic
coherence times, is relevant for liquid-phase samples, and remains
effective over interaction regions extending across many laser wavelengths,
even in the presence of dispersion.  We show how, by 3D shaping the
field's polarisation over the interaction region, we can achieve enantioselective
coherent control over electronic population transfer. Our ab-initio
simulations in the chiral molecule carvone yield a selectivity of
$\sim$30\% in the populations of the first excited electronic state,
three orders-of-magnitude higher than what is possible with circularly
polarised light ($\sim$0.01\%). These results bring all-optical enantioselective
photochemistry into the realm of practical applications. 
\end{abstract}
\maketitle

\renewcommand{\thefootnote}{$\dagger$}
\footnotetext{Present address: Department of Physics, Freie Universit\"at Berlin, 14195 Berlin, Germany.}
\renewcommand{\thefootnote}{$\ddagger$}
\footnotetext{Present address: Department of Chemistry, Imperial College London, W12 0BZ London, UK.}
\renewcommand{\thefootnote}{\arabic{footnote}}

One of the most powerful concepts in physics is the realisation that
different systems have different energy spectra. We rely on this not
only to distinguish different systems, but also to selectively excite
them, with far reaching consequences such as fluorescence microscopy
\citep{hellNobelLectureNanoscopy2015,betzigNobelLectureSingle2015,moernerNobelLectureSinglemolecule2015}
and magnetic resonance imaging \citep{lauterburAllScienceInterdisciplinary2005,mansfieldSnapshotMagneticResonance2004}.
However, when two systems differ only by a mirror reflection, like
opposite enantiomers (left- and right-handed versions of a chiral
molecule), they share the same spectra. Thus, enantioselective excitation
requires relying on light's properties beyond the photon energy, such
as the photon helicity. This imposes fundamental limitations on our
ability to manipulate chiral matter using light, with broad repercussions
throughout chemistry and biology.   

Beyond their well established role in the pharmaceutical and agrochemical
sectors, chiral molecules are emerging as biomarkers for diagnosing
diseases including cancer, Alzheimer, and diabetes \citep{liuDetectionAnalysisChiral2023}.
Moreover, molecular chirality is rapidly becoming an important asset
for nanotechnology \citep{brandtAddedValueSmallmolecule2017} in the
development of molecular motors \citep{feringaArtBuildingSmall2017}
and switches \citep{feringaMolecularSwitches2001}, spintronic devices
\citep{naamanChiralMoleculesElectron2019}, and self-assemblying structures
\citep{choBioinspiredChiralInorganic2023}. However, despite the
success and diversity of light-based approaches at distinguishing
opposite enantiomers \citep{ayusoUltrafastChiralityRoad2022a,rouxelMolecularChiralityIts2022,beginNonlinearHelicalDichroism2023,sparlingTwoDecadesImaging2025,koumarianouAssignmentfreeChiralityDetection2022,tutunnikovObservationPersistentOrientation2020,schwennickeEnantioselectiveTopologicalFrequency2022},
achieving highly enantioselective photo-excitation remains very challenging. 

For most chiral molecules, the interaction with circularly polarised
light leads to excited-state populations differing by less than 0.1\%
in opposite enantiomers \citep{l.greenfieldPathwaysIncreaseDissymmetry2021,raucciChiralPhotochemistryAchiral2022}.
 This difference stems from the interference between electric- and
magnetic-dipole interactions \citep{barronMolecularLightScattering2004,berovaComprehensiveChiropticalSpectroscopy2012}.
The difference is very small because the magnetic-dipole interaction
is usually much weaker than the electric-dipole interaction, and thus
their interference is only a small correction to the purely electric-dipole
contribution. The so-called superchiral fields aim to amend this
problem by considering structured electromagnetic field configurations
displaying regions where the ratio of the magnetic- to the electric-field
strength is enhanced \citep{tangEnhancedEnantioselectivityExcitation2011}.
However, these regions occur at minima of the electric field intensity
and are of deep-sub-wavelength scale, which poses important hurdles
for applications \citep{heDissymmetryEnhancementEnantioselective2018}. 

Multiphoton processes offer the opportunity to achieve enantioselective
population transfer in the absence of magnetic interactions altogether.
Instead of relying on (imbalanced) interference between electric-
and magnetic-dipole transitions, a multicolour field can drive an
enantiosensitive interference between two (balanced) electric-dipole
pathways. This is the essence of the coherent control approach to
enantioselective population transfer \citep{shapiroQuantumControlMolecular2012,gerbasiTheoryTwoStep2004,leibscherFullQuantumControl2022},
which has been recently implemented (largely independently from previous
works) in pioneering experiments using microwaves in the context of
rotational transitions in cold ($\sim$1 K) gas samples \citep{eibenbergerEnantiomerSpecificStateTransfer2017,perezCoherentEnantiomerSelectivePopulation2017,perezStateSpecificEnrichmentChiral2018,sunInducingTransientEnantiomeric2023}.
By velocity filtering the molecules, a resonant narrow-band ($<$1
MHz) UV laser can turn enantioselectivity in rotational states, to
enantioselectivity in electronic states, as beautifully shown in Refs.
\citep{leeQuantitativeStudyEnantiomerSpecific2022,leeNearcompleteChiralSelection2024}.
However, the rotational aspect of this technique and the use of narrow-band
UV lasers  limits it to the gas phase, hindering scaling up the number
of excited molecules for practical applications. Furthermore, the
narrow-band UV approach is ineffective on molecular species where
the excited state is short lived, as is often the case due to radiationless
relaxation \citep{schuurmanDynamicsConicalIntersections2018,borneUltrafastElectronicRelaxation2024}.
 Each of these fundamental limitations presents a major hurdle for
applications in photochemistry. 

Electronic excitation can trigger a broad range of photochemical phenomena
such as photoisomerization, photodissociation, and fluorescence. Thus,
driving such excitations with high chiral sensitivity naturally leads
to all-optical and highly enantioselective photochemistry. This is
in stark contrast to current approaches, where the enantioselectivity
stems entirely from intricate intermolecular interactions with tailored
molecules \citep{brimioulleEnantioselectiveCatalysisPhotochemical2015,tarafderChiralChromatographyMethod2021,genzinkChiralPhotocatalystStructures2022}.
All-optical approaches would bring distinct advantages with respect
to chemical methods in terms of generality, as well as in terms of
the temporal and spatial scales on which molecular chirality can be
manipulated.   

Here we show how to achieve highly enantioselective population transfer
between electronic states using current femtosecond laser technology
\citep{fattahiThirdgenerationFemtosecondTechnology2014,burgerCompactFlexibleHarmonic2017},
providing a route to enhancing the enantioselectivity of photochemical
reactions by three orders of magnitude.  Our approach relies on tailoring
a light field using only on an intense IR source and two of its harmonics.
It requires neither rotationally cold molecules, nor quadruply-resonant
excitation, nor narrow band excitation, nor long electronic coherence
times, can be extended to liquid samples \citep{brixnerPhotoselectiveAdaptiveFemtosecond2001},
and is effective throughout interaction regions extending over many
wavelengths, even in the presence of dispersion. Thus, we overcome
several fundamental limitations of previous proposals \citep{hokiSelectivePreparationEnantiomers2001,gerbasiTheoryTwoStep2004,perezStateSpecificEnrichmentChiral2018,yachmenevFieldInducedDiastereomersChiral2019,vitanovHighlyEfficientDetection2019,torosovEfficientRobustChiral2020,neufeldStrongChiralDichroism2021,leeQuantitativeStudyEnantiomerSpecific2022,leibscherFullQuantumControl2022,yePhasematchedLocallyChiral2023,sunInducingTransientEnantiomeric2023,leeNearcompleteChiralSelection2024}
and bring all-optical highly enantioselective photochemistry closer
to applications.

\subsection*{Results}

\begin{figure}
\begin{centering}
\includegraphics[width=0.95\textwidth]{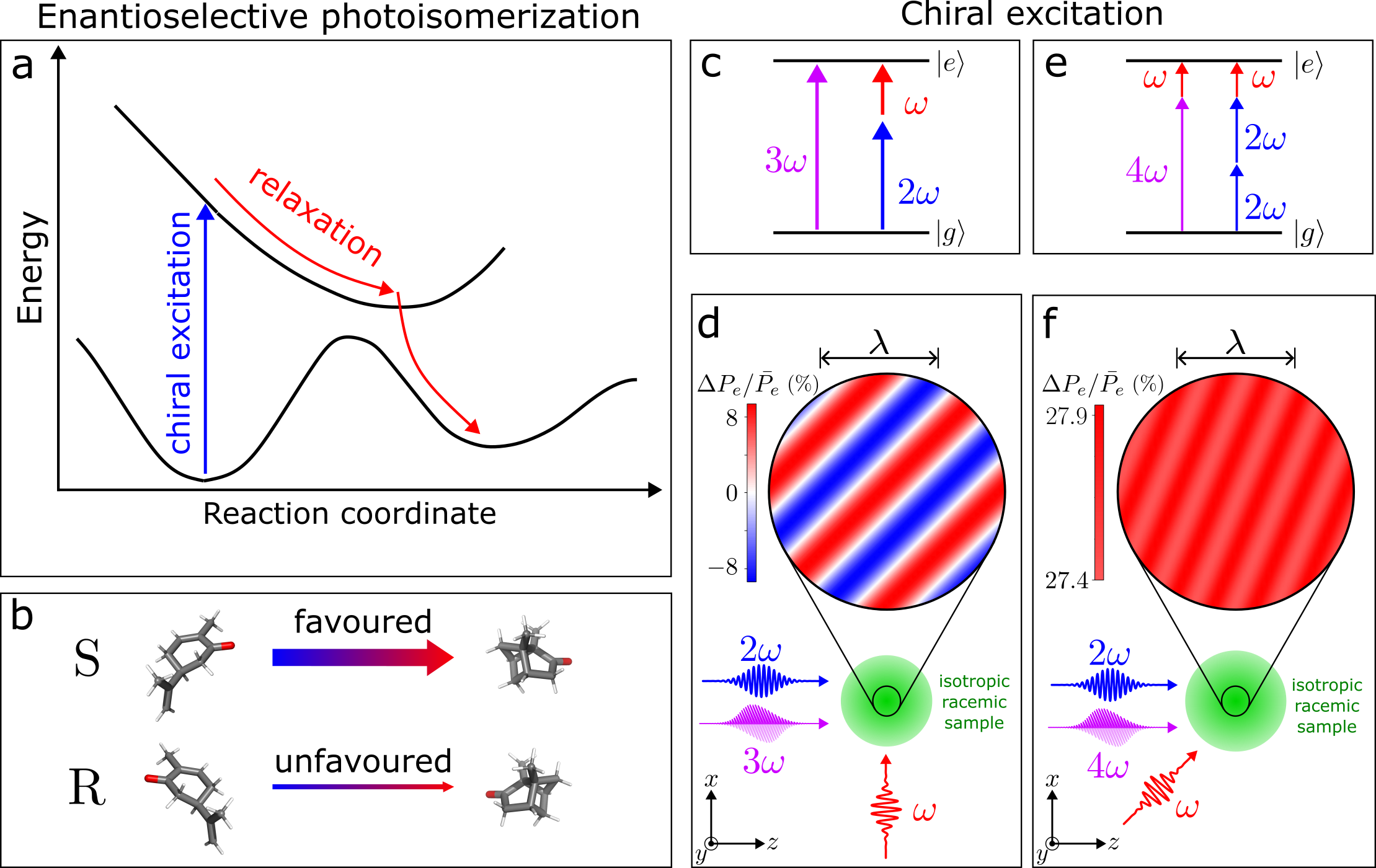}
\par\end{centering}
\caption{Highly enantioselective\textbf{ }photochemistry with nonlinear chiral
light-matter interactions. \textbf{a.} Scheme for enantioselective
photoisomerization with chiral light. Chiral light drives an electronic
excitation in a chiral molecule, which then relaxes to a new equilibrium
geometry. The potential energy surfaces determining the course of
the relaxation process are the same in both enantiomers. Chiral light-matter
interaction preferentially excites of one of the two enantiomers,
thus favouring its photoisomerization. \textbf{b.} In carvone, electronic
excitation triggers isomerization into carvonecamphor. Enantioselective
excitation favours the production of e.g. S carvonecamphor. \textbf{c.}
Enantioselective excitation of randomly oriented molecules can be
achieved in a 1- vs 2-photon coherent-control scheme, provided the
field's chirality $h^{\left(3\right)}$ {[}see Eq. (\ref{eq:h3}){]}
is non zero.\textbf{ d. }The chiral excitation in \textbf{c }can be
achieved with linearly polarised laser fields propagating as shown
($3\omega$ is polarised along $\hat{y}$). This approach leads to
spatial oscillations of $h^{\left(3\right)}$ occurring on a sub-wavelength
scale $(\lambda=2\pi c/\omega)$, which are directly reflected on
the enantioselectivity $\Delta P_{e}/\overline{P}_{e}=(P_{e}^{\mathrm{S}}-P_{e}^{\mathrm{R}})/[(P_{e}^{\mathrm{S}}+P_{e}^{\mathrm{R}})/2]$
(see inset), where $P_{e}^{\mathrm{X}}$ is the excited state population
of the $\mathrm{X}$ enantiomer after the interaction. The inset shows
results for Gaussian pulses with full width at half maximum $\mathrm{FWHM}=18\,\mathrm{fs}$
and peak intensities $I_{\omega}=I_{2\omega}=2.1\times10^{11}\,\mathrm{W/cm^{2}}$,
$I_{3\omega}=9.4\times10^{7}\,\mathrm{W/cm^{2}}$ \textbf{e.} Enantioselective
excitation can also be achieved in a 2- vs 3-photon coherent control
scheme, provided the field's chirality $h^{\left(5\right)}$ {[}see
Eq. (\ref{eq:h5}){]} is non zero. \textbf{f. }The chiral excitation
in \textbf{e} can be achieved with linearly polarised laser fields
propagating as shown ($4\omega$ is polarised along $\hat{y}$). In
the absence of dispersion, this configuration leads to an $h^{\left(5\right)}$
which is constant in space, resulting in virtually constant enantioselectivity
$\Delta P_{e}/\overline{P}_{e}$ across the interaction region (see
inset). The inset shows results for Gaussian pulses with $\mathrm{FWHM=22\,}\mathrm{fs}$,
$I_{\omega}=7.3\times10^{11}\,\mathrm{W/cm^{2}},$ $I_{2\omega}=9.4\times10^{11}\,\mathrm{W/cm^{2}}$,
$I_{4\omega}=1.6\times10^{10}\,\mathrm{W/cm^{2}}$. \label{fig:main}}
\end{figure}

We introduce our approach by considering two coherent control schemes
that involve a 1- vs 2-photon interference, and a 2- vs 3-photon interference.
We consider the enantioselective photoexcitation of an isotropic and
racemic sample of chiral molecules. That is, we assume that the molecules
are randomly oriented, an equal amount of the two opposite enantiomers,
and our objective is to tailor the light to preferentially excite
one enantiomer over the other. More precisely, we denote the orientation-averaged
populations in the first electronic excited state $|e\rangle$ of
the left (S) and right (R) enantiomers as $P_{e}^{\mathrm{S}}$ and
$P_{e}^{\mathrm{R}}$, respectively, and we aim to maximise the enantioselectivity
as quantified by $\Delta P_{e}/\overline{P}_{e}$, where $\Delta P_{e}\equiv P_{e}^{\mathrm{S}}-P_{e}^{\mathrm{R}}$
and $\overline{P}_{e}\equiv(P_{e}^{\mathrm{S}}+P_{e}^{\mathrm{R}})/2$.
First we present the proposed laser setup and the corresponding simulations,
and then we explain the analytical model that can be used to rationalise
the results. 

We illustrate our approach in carvone, a garden-variety chiral molecule
that played a central role in the early days of photochemistry \citep{ciamicianChemischeLichtwirkungen1908,schonbergPreparativeOrganicPhotochemistry1968,crimminsEnoneOlefinPhotochemical2004,malatestaLaserinducedCycloadditionsCarvone1982,tsipiIntramolecularPhotocycloadditionCarvone1987,zandomeneghiLaserPhotochemistryIntramolecular1980}.
Its S form can be found in caraway seeds and its R form in spearmint.
Photoexcitation of carvone with visible/UV light triggers its isomerization
(see Fig \ref{fig:main}a), a common process in organic molecules.
The photoisomerization of carvone into carvonecamphor \citep{ciamicianChemischeLichtwirkungen1908,buchiPhotochemicalReactionsVII1957,zandomeneghiLaserPhotochemistryIntramolecular1980,brackmannPhotocyclizationCarvoneCarvone1982,malatestaLaserinducedCycloadditionsCarvone1982,tsipiIntramolecularPhotocycloadditionCarvone1987}
was the first example of a {[}2+2{]} photocycloaddition, a prominent
photochemical reaction \citep{brimioulleEnantioselectiveLewisAcid2013,sarkarPhotochemicalCycloadditionOrganic2020}.
Once excited, both carvone enantiomers are equally likely to isomerise
into their respective form of carvonecamphor. Therefore, by making
the photoexcitation step enantioselective, i.e. by making $\Delta P_{e}\neq0$,
one can preferentially trigger one of the two mirror image photochemical
reactions (see Fig . \ref{fig:main}b).

\textbf{}Figures \ref{fig:main}c and d show a 1- vs 2-photon coherent
control scheme leading to highly enantioselective excitation. The
three frequencies are linearly polarised perpendicular to each other,
with the smaller frequency propagating at right angles to the other
two, as in the pioneering microwave experiments that demonstrated
enantioselective population transfer between rotational states \citep{eibenbergerEnantiomerSpecificStateTransfer2017,perezCoherentEnantiomerSelectivePopulation2017}.
Unlike in those works, the two-photon pathway does not rely on an
intermediate resonance, we use femtosecond pulses, and we focus on
electronic excitation in the perturbative regime, where two-pathway
coherent control of polyatomic molecules is well established \citep{wangPhaseControlAbsorption1996}.
We use frequencies $\omega$, $2\omega$, $3\omega$,  with $3\omega=3.89\,\mathrm{eV}$
resonant with the first electronic excitation.  Such phase-locked
harmonics can be generated from a single coherent source \citep{boydNonlinearOptics2008,burgerCompactFlexibleHarmonic2017}.
By adjusting the relative laser intensities we can balance the weights
of the two interfering pathways in Fig. \ref{fig:main}c. We keep
the pulses relatively short ($\sim$$20$ fs) to limit population
transfer and stay within the few-photon picture of Fig. \ref{fig:main}c,
as well as to avoid significant nuclear dynamics, but long enough
to avoid few-cycle effects.   We find that, since the fields propagate
at an angle to each other, their relative phase varies as a function
of position across the interaction region. This leads to a spatial
oscillation of the interference between the 1- and 2-photon pathways,
 resulting in sub-wavelength ($\lambda=2\pi c/\omega$$=956\,\mathrm{nm}$
in vacuum) oscillations of $\Delta P_{e}/\overline{P}_{e}$ between
$\pm9.2\%$, as captured by our ab-initio simulations (see Methods)
shown in the inset of Fig. \ref{fig:main}d. That is, the S molecules
are preferentially excited in the red regions of Fig. \ref{fig:main}d,
while the R molecules are preferentially excited in the blue regions,
and the centres of blue and red regions are less than a wavelength
apart. 

The sub-wavelength oscillation of $\Delta P_{e}/\overline{P}_{e}$
is a crucial factor when setting up future experiments with optical
wavelengths. Indeed, if the measurement spatially averages over red
and blue regions, the enantioselectivity will be washed out. Avoiding
this requires implementing special procedures, such as confining the
interaction region to only red or only blue regions, e.g. by delivering
the molecules via a flat liquid micro-jet \citep{galinisMicrometerthicknessLiquidSheet2017,luuExtremeUltravioletHigh2018,barnardDeliveryStableUltrathin2022,ferchaudInteractionIntenseFewcycle2022}
of sub-wavelength thickness and appropriate orientation. As we show
in the next subsection, by adjusting the beam geometry, it is possible
to significantly increase the spatial period of $\Delta P_{e}/\overline{P}_{e}$.
However, this comes at the cost of reducing the magnitude of $\Delta P_{e}/\overline{P}_{e}$.

\begin{figure}
\begin{centering}
\includegraphics[width=0.4\textwidth]{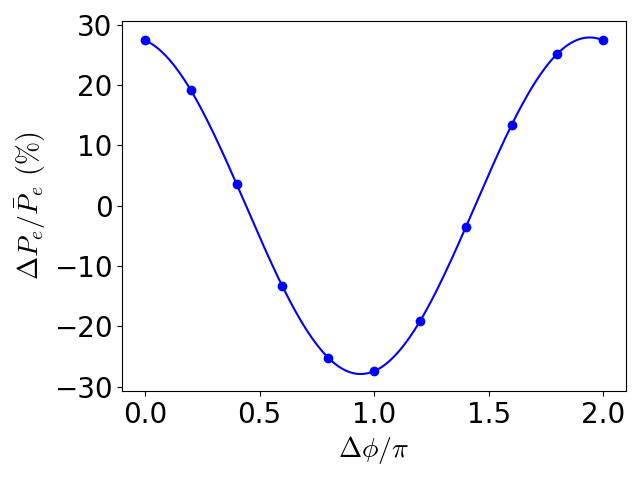}
\par\end{centering}
\caption{Coherent control of $\Delta P_{e}/\overline{P}_{e}$ as a function
of the the relative phase $\Delta\phi$ between $2\omega$ and $4\omega$
for the scheme shown in Fig. \ref{fig:main}f at a fixed point in
space. The dots show calculations and the solid line a fit to $A\cos(\Delta\phi+\delta)$
with $A=27.9\,\%$ and $\delta=0.186\,\mathrm{rad}$.  \label{fig:Coherent-control}}

\end{figure}

\textbf{} We can avoid the sub-wavelength oscillation of $\Delta P_{e}/\overline{P}_{e}$
by eliminating the dependence of the two-quantum-pathway interference
on the relative phase between frequencies propagating non-collinearly.
We achieve this by ensuring that the non-collinear $\omega$ contributes
one photon to each excitation pathway, as shown in the 2- vs 3-photon
scheme in Fig. \ref{fig:main}e. This leads to an exact cancellation
of the phase of $\omega$ in the interference term, which now only
depends on the relative phase $\Delta\phi=2\phi_{2\omega}-\phi_{4\omega}$
between the $2\omega$ and $4\omega$ beams (the control parameter).
Interestingly, our analytical and numerical results reveal that for
this excitation scheme, having the three polarisations perpendicular
to each other yields $\Delta P_{e}/\overline{P}_{e}=0$, in stark
contrast to the 1- vs 2-photon scheme. Here we choose the polarisations
and propagation directions as as depicted in Fig. \ref{fig:main}f,
with $\omega+4\omega$ resonant with the first electronic excitation
($\lambda=2\pi c/\omega=1.59$ $\mu$m). The inset of Fig. \ref{fig:main}f
shows that $\Delta P_{e}/\overline{P}_{e}$ barely oscillates around
the value $27.7\%$. The small oscillation can be attributed to small
contributions from other excitation pathways. Changing the relative
phase $\Delta\phi$ by $\pi$ leads to $\Delta P_{e}/\overline{P}_{e}=-27.9\%$,
i.e. preferential excitation of the opposite enantiomer. The control
of $\Delta P_{e}/\overline{P}_{e}$ as a function of $\Delta\phi$
is shown in Fig. \ref{fig:Coherent-control} at a fixed position in
space. For any $\Delta\phi$, the position dependence of $\Delta P_{e}/\overline{P}_{e}$
is similar to that shown in Fig. \ref{fig:main}f. Thus, our calculations
show that by careful design of the excitation scheme, enantioselectivity
can be coherently controlled and maintained virtually constant over
distances much greater than the wavelength.

\subsubsection*{Analytical model of enantioselective electronic population transfer}

\textbf{1- vs 2-photon scheme. } We write the electric fields as
$\tilde{\vec{E}}_{j}(\vec{r},t)=\mathrm{Re}\{\vec{E}_{j}(\vec{r})e^{-i\omega_{j}t}\}$,
$j=1,2,3$, where the complex vector $\vec{E}_{j}(\vec{r})$ encodes
the polarisation, amplitude, and phase associated to frequency $\omega_{j}$
at position $\vec{r}$.  The difference $\Delta P_{e}\equiv P_{e}^{\mathrm{R}}-P_{e}^{\mathrm{S}}$,
results from the interference between the two quantum pathways in
Fig. \ref{fig:main}c and is given by (see Methods)

\begin{equation}
\Delta P_{e}(\vec{r})=\mathrm{Re}\left\{ g^{(3)}h^{(3)}(\vec{r})\right\} ,\label{eq:DeltaP_3}
\end{equation}

where $g^{(3)}$ is a complex pseudoscalar with opposite signs for
opposite enantiomers recording the electronic transition dipoles and
detunings, and $h^{(3)}$ is the third-order chiral correlation function
characterising the field's chirality relevant for this particular
excitation \citep{ayusoSyntheticChiralLight2019}, 

\begin{equation}
h^{(3)}(\vec{r})\equiv\vec{E}_{3}^{*}\cdot\left(\vec{E}_{1}\times\vec{E}_{2}\right),\label{eq:h3}
\end{equation}

where $\vec{E}_{1}(\vec{r})$ and $\vec{E}_{2}(\vec{r})$ drive the
two-photon pathway, while $\vec{E}_{3}(\vec{r})$ drives the one-photon
pathway (see Fig. \ref{fig:main}c). Equations (\ref{eq:DeltaP_3})
and (\ref{eq:h3}) show that we can favour the excitation of either
S ($\Delta P_{e}>0$) or R ($\Delta P_{e}<0$) by controlling $h^{(3)}(\vec{r})$.
Indeed, $h^{(3)}$ is analogous to the helicity $h=\vec{E}^{*}\cdot\vec{B}$
emerging in one-photon circular dichroism, with $\vec{E}_{1}\times\vec{E}_{2}$
acting as an effective magnetic field $\vec{B}$, which is consistent
with replacing the magnetic field transition by a two-photon electric-dipole
transition.

Equation (\ref{eq:h3}) provides a recipe to maximise $|h^{(3)}|$,
and thus $\Delta P_{e}$, via polarisation shaping. One option is
to have the three polarisations linear and perpendicular to each other,
as in Fig. \ref{fig:main}d. An equally effective choice is to take
$\vec{E}_{2}$ and $\vec{E}_{3}$ circularly polarised and co-rotating,
and $\vec{E}_{1}$ linearly polarised perpendicular to the plane of
circular polarisation (e.g. $\vec{E}_{1}\propto\hat{z}$, $\vec{E}_{2}\propto\vec{E}_{3}\propto\hat{x}+i\hat{y}$).
Or $\vec{E}_{1}$ and $\vec{E}_{2}$ circularly polarised and counter-rotating,
and $\vec{E}_{3}$ linearly polarised and perpendicular to the plane
of circular polarisation (e.g. $\vec{E}_{1}\propto\hat{x}+i\hat{y}$,
$\vec{E}_{2}\propto\hat{x}-i\hat{y}$, $\vec{E}_{3}\propto\hat{z}$).
All of these choices maximise the triple product in Eq. (\ref{eq:h3}).

For plane-wave fields $\vec{E}_{j}(\vec{r})=|E_{j}|e^{i(\vec{k}_{j}\cdot\vec{r}-\phi_{j})}\hat{\epsilon}_{j}$,
we obtain $h^{(3)}(\vec{r})=\eta^{\left(3\right)}e^{i\left(\Delta\vec{k}\cdot\vec{r}-\Delta\phi\right)}$,
with $\eta^{\left(3\right)}\equiv|E_{1}E_{2}E_{3}|[\hat{\epsilon}_{3}^{*}\cdot(\hat{\epsilon}_{1}\times\hat{\epsilon}_{2})]$,
and $\Delta P_{e}(\vec{r})$ becomes a plane standing wave,

\begin{equation}
\Delta P_{e}(\vec{r})=|g^{(3)}\eta^{(3)}|\cos(\Delta\vec{k}\cdot\vec{r}-\Delta\phi+\delta),\label{eq:deltaP_wave}
\end{equation}

where $\Delta\vec{k}\equiv\vec{k}_{1}+\vec{k}_{2}-\vec{k}_{3}$, $\Delta\phi\equiv\phi_{1}+\phi_{2}-\phi_{3}$,
and $\delta\equiv\mathrm{arg}(g^{\left(3\right)}\eta^{\left(3\right)})$.
Thus, $\Delta P_{e}(\vec{r})$ oscillates in space with a period $\Lambda\equiv2\pi/|\Delta\vec{k}|$,
an amplitude determined by the strength of the chiral light-matter
coupling $|g^{(3)}\eta^{(3)}|\propto|\hat{\epsilon}_{3}^{*}\cdot(\hat{\epsilon}_{1}\times\hat{\epsilon}_{2})|$,
and a phase controlled by the phase between the two pathways $\Delta\phi$
and the light-matter coupling phase $\delta$. The magnitude $|\hat{\epsilon}_{3}^{*}\cdot(\hat{\epsilon}_{1}\times\hat{\epsilon}_{2})|$
determines the degree of enantioselectivity, while the `chiral coherence
length' $\Lambda/2=\pi/|\Delta\vec{k}|$ determines the distance over
which enantioselectivity remains `in phase' and is not washed out
by spatial averaging. Thus, it determines the maximum size of the
interaction region along $\Delta\vec{k}$'s direction.  

Equation (\ref{eq:deltaP_wave}) provides a recipe for maximising
the enantioselectivity $\Delta P_{e}$ of the photochemical excitation
via light shaping. Ideally, one would like to maximise both the degree
of enantioselectivity and the chiral coherence length. Unfortunately,
the field polarisation that maximises $|\hat{\epsilon}_{3}^{*}\cdot(\hat{\epsilon}_{1}\times\hat{\epsilon}_{2})|$
minimises $\Lambda$, and vice versa. Indeed, achieving a non-zero
$|\hat{\epsilon}_{3}^{*}\cdot(\hat{\epsilon}_{1}\times\hat{\epsilon}_{2})|$
with radiation (transverse) fields requires at least one frequency
propagating at an angle to the other two, which generally leads to
$|\Delta\vec{k}|\neq0$. For example, in Fig. \ref{fig:main}d, the
wave vectors $\vec{k}_{1}=\omega\hat{x}/c$, $\vec{k}_{2}=2\omega\hat{z}/c$,
and $\vec{k}_{3}=3\omega\hat{z}/c$, result in $\Delta\vec{k}=\omega(\hat{x}-\hat{z})/c$,
which explains the resulting direction and periodicity ($\Lambda=\lambda/\sqrt{2}$,
where $\lambda=2\pi c/\omega$) of the oscillations of $\Delta P_{e}/\overline{P}_{e}$.
Finding the right balance between $|\hat{\epsilon}_{3}^{*}\cdot(\hat{\epsilon}_{1}\times\hat{\epsilon}_{2})|$
and $\Lambda$ requires careful consideration the particular experimental
conditions. 

\begin{figure}
\begin{centering}
\par\end{centering}
\begin{centering}
\includegraphics[width=0.5\textwidth]{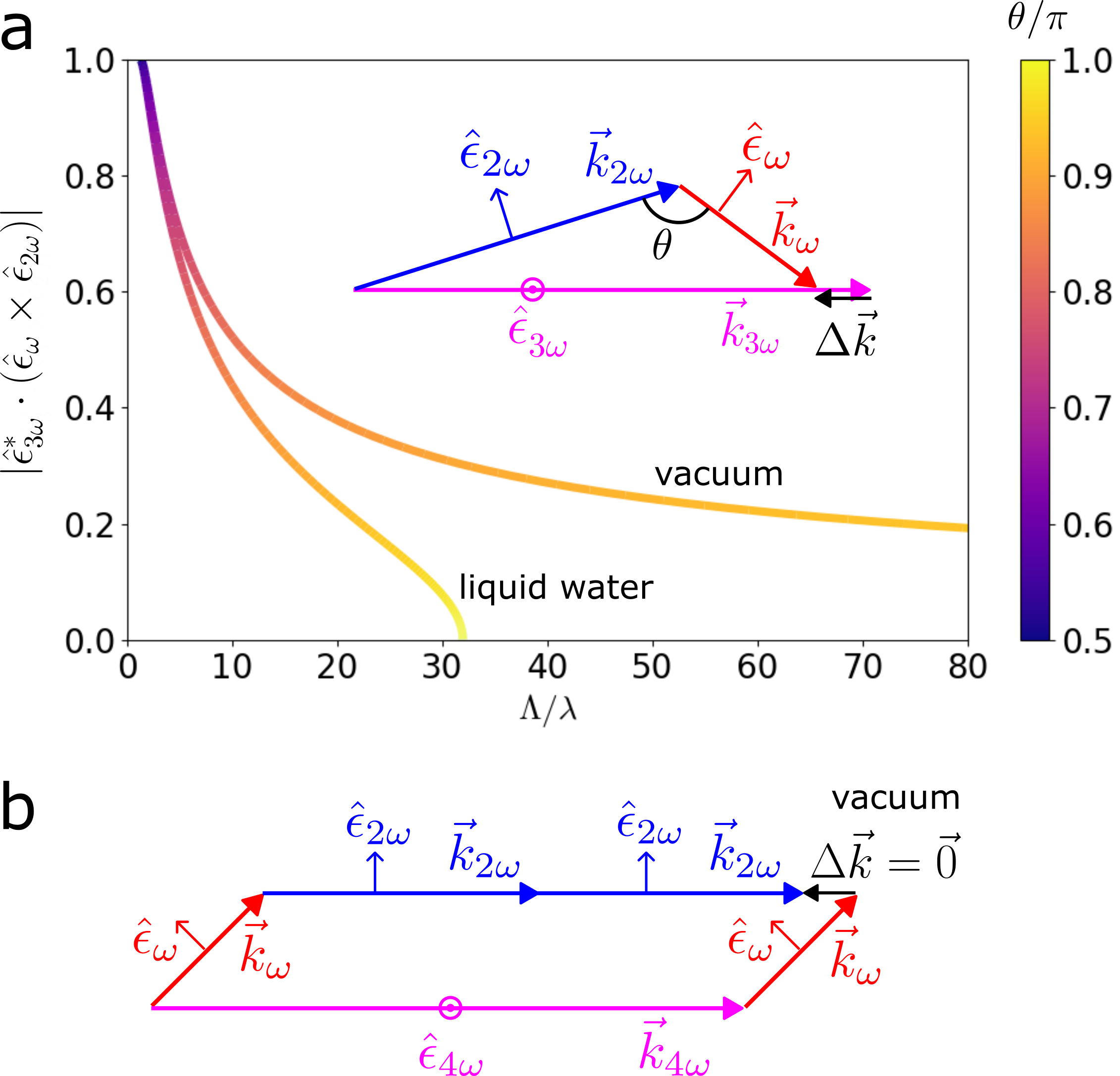}
\par\end{centering}
\caption{\textbf{a.} Trade-off between $|\hat{\epsilon}_{3\omega}^{*}\cdot(\hat{\epsilon}_{\omega}\times\hat{\epsilon}_{2\omega})|$
and $\Lambda/\lambda$ in a beam arrangement that attempts to maximise
both $|\hat{\epsilon}_{3\omega}^{*}\cdot(\hat{\epsilon}_{\omega}\times\hat{\epsilon}_{2\omega})|$
and $\Lambda=2\pi/|\Delta\vec{k}|$, which determine the amplitude
and spatial periodicity, respectively, of the enantiosensitive population
$\Delta P_{e}$ in the 1- vs 2-photon control scheme {[}see Eq. (\ref{eq:deltaP_wave}){]}.
The upper curve is in the absence of dispersion and the lower curve
assuming the dispersion of liquid water. $\lambda=2\pi c/\omega=956\,\mathrm{nm}$
in vacuum and $\lambda=2\pi c/n\omega=720\,\mathrm{nm}$ in liquid
water. $|\hat{\epsilon}_{3\omega}^{*}\cdot(\hat{\epsilon}_{\omega}\times\hat{\epsilon}_{2\omega})|=1$
when $\Lambda/\lambda=1.31$ in vacuum and $\Lambda/\lambda=1.26$
in liquid water.\textbf{ b. }There is no trade-off in the 2- vs 3-photon
control scheme chosen for the simulations in Fig. \ref{fig:main}f.
This geometry yields $\Lambda/\lambda\rightarrow\infty$ ($|\Delta\vec{k}|=0$)
in the absence of dispersion, and $\Lambda/\lambda=33$ assuming the
dispersion of liquid water, without incurring in reductions of the
amplitude of the enantioselective population $\Delta P_{e}$, determined
by $|[\hat{\epsilon}_{2\omega}\cdot(\hat{\epsilon}_{\omega}\times\hat{\epsilon}_{4\omega}^{*})](\hat{\epsilon}_{2\omega}\cdot\hat{\epsilon}_{\omega}^{*})|$
{[}see Eqs. (\ref{eq:h5}) and (\ref{eq:DeltaP_5_planeWave}){]}.
$\lambda=2\pi c/\omega=1.59\,\mu\mathrm{m}$ in vacuum and $\lambda=2\pi c/n\omega=1.21\,\mu\mathrm{m}$
in liquid water. \label{fig:tradeoff}}
\end{figure}

Let us illustrate the trade-off between the degree of enantioselectivity
and the chiral coherence length by considering the beam arrangement
in Fig. $\ref{fig:tradeoff}a$, where $\vec{k}_{1}$ and $\vec{k}_{2}$
are at a variable angle $\theta$ and their sum is set parallel to
$\vec{k}_{3}$ to minimise $|\Delta\vec{k}|$ (maximise $\Lambda$).
This arrangement provides the maximal $\Lambda$ for a given $|\hat{\epsilon}_{3}^{*}\cdot(\hat{\epsilon}_{1}\times\hat{\epsilon}_{2})|$.
A simple geometrical argument shows that $|\hat{\epsilon}_{3}^{*}\cdot(\hat{\epsilon}_{1}\times\hat{\epsilon}_{2})|=\sin\theta$
and $\Lambda=2\pi(k_{3}-\sqrt{k_{1}^{2}+k_{2}^{2}-2k_{1}k_{2}\cos\theta})^{-1}$.
The curves shown in Fig. \ref{fig:tradeoff}a are for propagation
in the absence of dispersion (vacuum, upper curve) and in a typical
dispersive medium (liquid water, lower curve), for $\omega_{1}=\omega$,
$\omega_{2}=2\omega$, $\omega_{3}=3\omega$.   Both curves show
that an increase in $\Lambda$ comes with a decrease in $|\hat{\epsilon}_{3}^{*}\cdot(\hat{\epsilon}_{1}\times\hat{\epsilon}_{2})|$.
For example, increasing $\Lambda/\lambda$ from $1$ to $10$ comes
at the cost of decreasing $|\hat{\epsilon}_{3}^{*}\cdot(\hat{\epsilon}_{1}\times\hat{\epsilon}_{2})|$
from 1 to 0.4 (0.5) in liquid water (vacuum). In liquid water, dispersion
brings $|\hat{\epsilon}_{3}^{*}\cdot(\hat{\epsilon}_{1}\times\hat{\epsilon}_{2})|$
to zero at $\Lambda/\lambda=32$, setting a hard limit on the maximum
value of $\Lambda$. In the absence of dispersion, the curve flattens
out for large values of $\Lambda/\lambda$, and e.g. $|\hat{\epsilon}_{3}^{*}\cdot(\hat{\epsilon}_{1}\times\hat{\epsilon}_{2})|=0.1$
at $\Lambda/\lambda=300$. Note that reduction of the maximum enantioselectivity
by a factor of 10 may still yield much higher enantioselectivity than
circularly polarised light. Therefore, while this scheme is not restricted
to sub-wavelength interaction regions, it is most effective for interaction
regions smaller than about $5\lambda$ in $\Delta\vec{k}$'s direction,
especially in dispersive media. This scale is within reach of current
liquid microjet technology \citep{galinisMicrometerthicknessLiquidSheet2017,luuExtremeUltravioletHigh2018,barnardDeliveryStableUltrathin2022,ferchaudInteractionIntenseFewcycle2022}.

\global\long\def\five#1#2#3#4#5{\left[#1\cdot\left(#2\times#3\right)\right]\left(#4\cdot#5\right)}%

\global\long\def\Five#1#2#3#4#5{\left(\begin{array}{c}
\five{#1}{#2}{#3}{#4}{#5}\\
\five{#1}{#2}{#4}{#3}{#5}\\
\five{#1}{#2}{#5}{#3}{#4}\\
\five{#1}{#3}{#4}{#2}{#5}\\
\five{#1}{#3}{#5}{#2}{#4}\\
\five{#1}{#4}{#5}{#2}{#3}
\end{array}\right)}%

\textbf{2- vs 3-photon scheme. }Applying the same theoretical framework,
we obtain (see Methods), 

\begin{equation}
\Delta P_{e}(\vec{r})=\mathrm{Re}\left\{ \vec{g}^{(5)}\cdot\vec{h}^{(5)}(\vec{r})\right\} ,\label{eq:DeltaP_5}
\end{equation}

where, in contrast to Eq. (\ref{eq:DeltaP_3}), $\vec{g}^{(5)}$ and
$\vec{h}^{(5)}$ are vectors (with pseudoscalar components). $\vec{h}^{(5)}$
encodes the field's handedness relevant for this excitation scheme,

\begin{equation}
\vec{h}^{\left(5\right)}(\vec{r})\equiv\left(\begin{array}{c}
[\vec{E}_{2}\cdot(\vec{E}_{1}\times\vec{E}_{1}^{*})](\vec{E}_{2}\cdot\vec{E}_{3}^{*})\\{}
[\vec{E}_{2}\cdot(\vec{E}_{1}\times\vec{E}_{3}^{*})](\vec{E}_{2}\cdot\vec{E}_{1}^{*})\\{}
[\vec{E}_{2}\cdot(\vec{E}_{1}^{*}\times\vec{E}_{3}^{*})](\vec{E}_{2}\cdot\vec{E}_{1})
\end{array}\right),\label{eq:h5}
\end{equation}

where $\vec{E}_{1}(\vec{r})$ contributes one photon to each pathway,
$\vec{E}_{2}(\vec{r})$ contributes 2 photons to one pathway, and
$\vec{E}_{3}(\vec{r})$ contributes one photon to the other pathway.
In Fig. \ref{fig:main}d, $\vec{E}_{1}$, $\vec{E}_{2}$, and $\vec{E}_{3}$
correspond to the fields at frequencies $\omega_{1}=\omega$, $\omega_{2}=2\omega$,
and $\omega_{3}=4\omega$, respectively. 

Having a vector of handedness, rather than a single number, reflects
that there is more than one independent way to construct a local chirality
measure with the five available electric field vectors  in the interference
term. Each component of $\vec{h}^{\left(5\right)}$ is more nuanced
than $h^{(3)}$, as they involve not only a triple product but also
a dot product. Thus, for a component of $\vec{h}^{\left(5\right)}$
to be non-zero, not only three of the electric field vectors must
be non-coplanar, the other two must be non-orthogonal. In particular,
if the three frequencies are linearly polarised perpendicular to each
other, all components of $\vec{h}^{(5)}$ vanish, leading to $\Delta P_{e}=0$,
in stark contrast to $h^{\left(3\right)}$. 

A fundamental consequence of the vectorial nature of $\vec{h}^{\left(5\right)}$
in Eq. (\ref{eq:DeltaP_5}) is that the polarisation dependence of
$\Delta P_{e}$ is molecule specific. In the case of $h^{\left(3\right)}$
and Eq. (\ref{eq:DeltaP_3}), the dependence of $\Delta P_{e}$ on
the polarisations of $\vec{E}_{1}$, $\vec{E}_{2}$, and $\vec{E}_{3}$
is the same for all molecules up to a relative phase, it varies as
$\vec{E}_{3}^{*}\cdot(\vec{E}_{1}\times\vec{E}_{2})$. However, in
the case of $\vec{h}^{\left(5\right)}$ and Eq. (\ref{eq:DeltaP_5}),
since each component of $\vec{h}^{\left(5\right)}$ is weighted by
a different component of $\vec{g}^{\left(5\right)}$, and each molecule
has a different distribution of $\vec{g}^{\left(5\right)}$ components,
the behaviour of $\Delta P_{e}$ as a function of the polarisations
of $\vec{E}_{1}$, $\vec{E}_{2}$ and $\vec{E}_{3}$ is molecule specific.
The polarisations used in Fig. \ref{fig:main}e, $\hat{\epsilon}_{1}=(\hat{x}-\hat{z})/\sqrt{2}$,
$\hat{\epsilon}_{2}=\hat{x}$, $\hat{\epsilon}_{3}=\hat{y}$, yield
$\vec{h}^{(5)}\propto(0,1,1)$. In contrast, choosing $\omega_{1}$
circularly polarised and $\omega_{2}$ and $\omega_{3}$ linearly
polarised perpendicular to the plane of circular polarisation yields
$\vec{h}^{(5)}\propto(1,0,0)$. Such polarisation combination has
been recently suggested in Ref. \citep{yePhasematchedLocallyChiral2023}.
Alternatively, choosing $\omega_{1}$ circularly polarised in the
arrangement of Fig. \ref{fig:main}e yields $\vec{h}^{(5)}\propto(0,-1,1)$.
The merit of each of these polarisation combinations and their corresponding
$\vec{h}^{(5)}$ vectors depends on the details of $\vec{g}^{(5)}$,
which varies from molecule to molecule and remains to be systematically
investigated. 

For plane-wave electric fields $\vec{E}_{j}(\vec{r})=|E_{j}|e^{i(\vec{k}_{j}\cdot\vec{r}-\phi_{j})}\hat{\epsilon}_{j}$,
we obtain $\vec{h}^{(5)}(\vec{r})=\vec{\eta}^{\left(5\right)}e^{i\left(\Delta\vec{k}\cdot\vec{r}-\Delta\phi\right)}$,
where $\vec{\eta}^{(5)}$ is obtained by replacing $\vec{E}_{j}$
by $|E_{j}|\hat{\epsilon}_{j}$ in Eq. (\ref{eq:h5}), and

\begin{equation}
\Delta P_{e}(\vec{r})=|\vec{g}^{(5)}\cdot\vec{\eta}^{(5)}|\cos(\Delta\vec{k}\cdot\vec{r}-\Delta\phi+\delta)\label{eq:DeltaP_5_planeWave}
\end{equation}

where $\Delta\vec{k}=2\vec{k}_{2}-\vec{k}_{3}$, $\Delta\phi=2\phi_{2}-\phi_{3}$,
$\delta=\mathrm{arg}(\vec{g}^{(5)}\cdot\vec{\eta}^{\left(5\right)})$.
In contrast to the 1- vs 2-photon case, here the $\omega_{1}$ photon
contributes to both pathways. Therefore, $\Delta P_{e}$ does not
depend on $\vec{k}_{1}$ and we can set $|\Delta\vec{k}|=0$ by having
co-propagating $\omega_{2}$ and $\omega_{3}$ (up to dispersion),
see Figs. \ref{fig:main}f and \ref{fig:tradeoff}b, without compromising
the degree of enantioselectivity, determined by $|\vec{g}^{(5)}\cdot\vec{\eta}^{(5)}|$.
Even when including dispersion, we still get $|\Delta\vec{k}|\ll k_{2}$,
leading to chiral coherence lengths $\Lambda/2\gg\lambda$. For example,
in liquid water, for the wavelengths used in Fig. \ref{fig:main}f,
$\Delta P_{e}$ would oscillate with a spatial period $\Lambda=2\pi/|\Delta\vec{k}|=33\lambda$,
where $\lambda=2\pi c/n\omega_{1}$.   Equation (\ref{eq:DeltaP_5_planeWave})
also shows that the preferential excitation of S ($\Delta P_{e}>0$)
or R ($\Delta P_{e}<0$) enantiomers can be controlled via the relative
phase $\Delta\phi=2\phi_{2}-\phi_{3}$ of the co-propagating frequencies,
in accordance with the numerical results in Fig. \ref{fig:Coherent-control}
showing up to 27.9\% enantioselectivity. 

\subsection*{Discussion}

Coherent control of electronic population in polyatomic molecules
has been experimentally demonstrated in gas and liquid samples \citep{wangPhaseControlAbsorption1996,brixnerQuantumControlGasPhase2003}.
Our work brings highly enantioselective population transfer (EPT)
in randomly oriented samples with current experimental capabilities
\citep{burgerCompactFlexibleHarmonic2017}. Indeed, state-of-the-art
field synthesis \citep{fattahiThirdgenerationFemtosecondTechnology2014,cirmiOpticalWaveformSynthesis2023}
can provide optical pulses carrying multiple frequencies spanning
from the UV to the near-IR, with high intensities, and controlled
relative phases, at kHz repetition rates \citep{muckeWaveformNonlinearOptics2015,burgerCompactFlexibleHarmonic2017,alismailMultioctaveCEPstableSource2020,xueCustomTailoredMultiTWOptical2021,alqattanAttosecondLightField2022,ridenteElectroopticCharacterizationSynthesized2022}.
Furthermore, the integration of thin liquid sheet technology in femtosecond
laser setups\citep{ferchaudInteractionIntenseFewcycle2022,barnardDeliveryStableUltrathin2022}
allows limiting the interaction region of liquid samples to a few
$\mu$m, which is the scale over which we have shown that enantioselectivity
can be maintained on the order of 10\% in 1- vs 2-photon coherent
control (Figs. \ref{fig:main}d and \ref{fig:tradeoff}a). This degree
of enantioselectivity is three orders of magnitude above that found
with circularly polarised light for the same transition \citep{lambertOpticalActivityCarvone2012}.
It also outperforms superchiral light, which enhances enantioselectivity
by one order of magnitude with respect to circularly polarised light,
over regions limited to $\sim$$10$ nm \citep{tangEnhancedEnantioselectivityExcitation2011,heDissymmetryEnhancementEnantioselective2018}.
Our 2- vs 3-photon scheme leads to enantioselectivity close to 30\%,
maintained throughout tens of $\mu\mathrm{m}$, and merely requires
a stable relative phase between the two co-propagating frequencies.
Furthermore, we do not rely on intermediate resonances to achieve
these results, which means that (i) unwanted processes resulting from
excitation of other states are avoided and (ii) the same setup can
be easily adjusted to address a wide range of molecular species. 

Our analytical theory reveals the physical mechanism underlying EPT
by identifying the relevant molecular and field pseudoscalars, as
well as the role of polarisation and propagation for two off-resonant
multiphoton excitation schemes. It also provides a simple recipe for
controlling and maximising the degree of enantioselectivity via light
shaping, according to the specific properties of the molecular sample
and the capabilities of each laboratory. The vectorial character
of $\vec{h}^{(5)}$ leads to molecule-specific optimisation of the
enantioselectivity that goes beyond simply relying on differences
in energy-level structure. This offers interesting perspectives for
selectively addressing a particular chiral species in mixtures containing
molecules with similar spectra. Such mixtures are often unavoidable,
as even pure samples of polyatomic species typically exist as a mixture
of several conformers with similar energy-level structure.

There are several alternatives to probe EPT to an electronic state.
In the case where the excited state decays via isomerization, as in
carvone, EPT will favour the creation of one of the two enantiomers
of the photoproduct. Thus, starting from a racemic sample, EPT will
lead to a non-racemic mixture, both for starting and product molecules
\citep{balavoinePreparationChiralCompounds1974,meinertPhotonenergyControlledSymmetryBreaking2014}.
This change can be monitored using standard enantiosensitive methods,
either spectroscopically or chemically, as usually done in asymmetric
catalysis experiments \citep{bauerCatalyticEnantioselectiveReactions2005,genzinkChiralPhotocatalystStructures2022}.
Alternatively, if the excited state decays via fluorescence, one could
start with an enantiopure sample and monitor how fluorescence changes
upon reversal of the field chirality ($h^{(3)}$ or $\vec{h}^{(5)}$),
as done with circularly polarised light or superchiral light \citep{tangEnhancedEnantioselectivityExcitation2011}.
Analogously, in the gas phase, instead of measuring fluorescence,
one could also probe the excited state population via photoelectron
spectroscopy \citep{wanieCapturingElectrondrivenChiral2024a}, and
monitor ionisation yields as a function of $h^{(3)}$ or $\vec{h}^{(5)}$.

Our all-optical femtosecond-laser approach to EPT between electronic
states naturally leads to highly enantioselective photochemistry,
which offers unique opportunities for chemical synthesis \citep{genzinkChiralPhotocatalystStructures2022}.
And since our approach does not rely on chiral sensitisers, it circumvents
limitations and challenges intrinsic to intermolecular interactions,
such as static enantioselectivity, synthesis of custom chiral sensitisers,
and intricate reaction mechanisms. Thus, it offers fundamentally new
opportunities, such as rapid switching of enantioselectivity (via
phase control). This is a key capability required for the implementation
of chiroptical molecular switches \citep{feringaMolecularSwitches2001,huckDynamicControlAmplification1996},
which are hindered by the small enantioselectivity obtained with circularly
polarised light.

\textcolor{red}{}

\subsection*{Methods}

\subsubsection*{Simulations}

In our simulations, we use the fixed nuclei approximation and the
electric-dipole approximation. For a given set of electric field parameters,
a fixed molecular orientation, and a fixed point in the interaction
region, we expand the electronic wave function in terms of the unperturbed
eigenstates and solve the TDSE for the expansion coefficients. The
population of the first excited state is then averaged over molecular
orientations. We also scan this population as a function of the relative
phase between the different frequencies of the electric field. This
scan is used to reconstruct the population as a function of molecular
position across the interaction region. This reconstruction is exact
in the limit of long pulses with planar wave fronts. 

We perform electronic structure calculations to compute the transition
energies and dipole moments required to solve the TDSE. The structure
of the most stable conformer of R carvone was taken from Ref. \citep{lambertOpticalActivityCarvone2012}
and reoptimized in the current work for consistency using Density
Functional Theory (DFT) \citep{kohnDensityFunctionalTheory1996,zieglerApproximateDensityFunctional2002}
with the B3LYP \citep{leeDevelopmentColleSalvettiCorrelationenergy1988,beckeDensityfunctionalExchangeenergyApproximation1988}
functional and the 6-311G(d,p) \citep{mcleanContractedGaussianBasis2008}
basis set available in Gaussian 16 \citep{g16}. The excitation energies
and transition moments were evaluated for the first 100 excited states
using time-dependent DFT with the CAM-B3LYP functional and the d-aug-cc-pVDZ
basis set. The transition moments between all excited states were
computed with the Multiwfn software \citep{luMultiwfnMultifunctionalWavefunction2012}.
The performance of the CAM-B3LYP functional in the calculation of
spectroscopic properties of chiral systems has been discussed in detail
in Refs. \citep{toroTwoPhotonAbsorptionCircular2010,toroTwophotonAbsorptionCircularlinear2010,linVibronicallyResolvedElectronic2008,linTheoryVibrationallyResolved2009,rizzoInitioStudyExcited2011}
and has been recently shown to predict circular dichroism spectra
in good agreement with results obtained from highly accurate EOM-CCSD
calculations \citep{andersenProbingMolecularChirality2022}.

The integration over orientations is carried out using a 14-point
Lebedev grid for $\alpha$ and $\beta$, and a trapezoidal rule with
9 equally spaced points in $\gamma$, for a total of 126 orientations,
where $\alpha,\beta,\gamma$ are Euler angles in the $zyz$ convention.
Increasing the number of orientations to 286 (26-point Lebedev in
$\alpha$ and $\beta$ and 11 in $\gamma$) did not lead to any significant
change.

For an electric field consisting of several frequencies $\{\omega_{j}\}_{j=1}^{N}$,
each of them a plane wave propagating along the unit vector $\hat{k}_{j}$
with speed $c_{j}$ and phase $\phi_{j}$, we can write 

\begin{align}
\vec{E}\left(\vec{r},t;\phi_{1},\phi_{2},\dots,\phi_{N}\right) & \equiv\sum_{j=1}^{N}E_{j}e^{-i\left(\omega_{j}t+\phi_{j}-\frac{\omega_{j}}{c_{j}}\hat{k}_{j}\cdot\vec{r}\right)}\hat{\epsilon}_{j}.
\end{align}

Taking the frequency $\omega_{1}$ as a reference, one can show that
a shift in position $\Delta\vec{r}$ is equivalent to a simultaneous
shift of time and of the phases of the rest of frequencies $\Delta\phi_{j}$,
$j=2,\dots N,$ according to 

\begin{equation}
\vec{E}\left(\vec{r}+\Delta\vec{r},t;\phi_{1},\phi_{2},\dots,\phi_{N}\right)=\vec{E}\left(\vec{r};t+\Delta t;\phi_{1},\phi_{2}+\Delta\phi_{2},\dots,\phi_{N}+\Delta\phi_{N}\right),\label{eq:E(r+d)}
\end{equation}

where 

\begin{equation}
\Delta t=-\frac{\hat{k}_{1}\cdot\Delta\vec{r}}{c_{1}},\qquad\Delta\phi_{j}=\omega_{j}\left(\frac{1}{c_{1}}\hat{k}_{1}-\frac{1}{c_{j}}\hat{k}_{j}\right)\cdot\Delta\vec{r}.
\end{equation}

An overall time shift of the electric field does not change the final
state populations. But the final state populations do depend on the
phases $\phi_{j}$. Thus, knowing the populations for all $\phi_{j}$
at a given position $\vec{r}$ is enough to recover the populations
at any other point $\vec{r}+\Delta\vec{r}$ via Eq. (\ref{eq:E(r+d)}).
This approximation is valid as long as the temporal envelopes do not
change significantly upon the shift by $\Delta t$. That is, if the
pulse envelopes change on a time scale $\tau$, Eq. (\ref{eq:E(r+d)})
is valid provided $|\Delta\vec{r}|\ll c_{1}\tau$. Taking $c_{1}$
equal to the speed of light in vacuum and $\tau=100\,\mathrm{fs}$,
this yields $|\Delta\vec{r}|\ll30\,\mu\mathrm{m}$. In our simulations,
we use relatively short pulses $\tau\approx20\,\mathrm{fs}$ for the
sake of reducing the computing time and avoiding nuclear dynamics,
and thus our reconstruction is strictly valid only for $|\Delta\vec{r}|\ll6\,\mu\mathrm{m}$.
However, our approach can be trivially extended to longer pulses and
thus bigger $|\Delta\vec{r}|$. 

\subsubsection*{Analytical model for 1- vs 2-photon isotropic chiral coherent control}

Consider a chiral molecule interacting with a three-colour field given
by $\tilde{\vec{E}}(t)=\sum_{j=1}^{3}\tilde{\vec{E}}_{j}(t)$, where
$\tilde{\vec{E}}_{j}(t)=\mathcal{E}_{j}(t)\mathrm{Re}\{\vec{E}_{j}e^{-i\omega_{j}t}\}$,
$\mathcal{E}_{j}(t)$ describes a smooth temporal envelope and the
complex vector $\vec{E}_{j}$ encodes the amplitude, polarisation,
and phase of the frequency $\omega_{j}$. The frequencies are such
that $\omega_{1}+\omega_{2}=\omega_{3}$ is resonant with the transition
from the ground state $|g\rangle$ to the excited state $|e\rangle$.
This leads to three sets of excitation pathways: a one photon pathway
$|g\rangle\overset{\omega_{3}}{\rightarrow}|e\rangle$, and two sets
of two-photon pathways $|g\rangle\overset{\omega_{2}}{\rightarrow}|n\rangle\overset{\omega_{1}}{\rightarrow}|e\rangle$
and $|g\rangle\overset{\omega_{1}}{\rightarrow}|n\rangle\overset{\omega_{2}}{\rightarrow}|e\rangle$,
that differ only by the photon ordering, and where $|n\rangle$ is
an intermediate state that has to be summed over. The probability
amplitude for state $|e\rangle$ after the interaction is given by

\begin{equation}
a_{e}=a_{e}^{\left(\omega_{3}\right)}+\sum_{n}\left(a_{e,n}^{\left(\omega_{1},\omega_{2}\right)}+a_{e,n}^{\left(\omega_{2},\omega_{1}\right)}\right),
\end{equation}

where perturbation theory yields 

\begin{equation}
a_{e}^{(\omega_{3})}=A^{\left(\omega_{3}\right)}\left(\vec{d}_{e,g}\cdot\vec{E}_{3}\right),
\end{equation}

\begin{equation}
a_{e,n}^{\left(\omega_{1},\omega_{2}\right)}=A_{n}^{\left(\omega_{1},\omega_{2}\right)}\left(\vec{d}_{e,n}\cdot\vec{E}_{1}\right)\left(\vec{d}_{n,g}\cdot\vec{E}_{2}\right),
\end{equation}

and $a_{e,n}^{\left(\omega_{2},\omega_{1}\right)}$ is given by $a_{e,n}^{\left(\omega_{1},\omega_{2}\right)}$
after replacing $A_{n}^{\left(\omega_{1},\omega_{2}\right)}$ by $A_{n}^{\left(\omega_{2},\omega_{1}\right)}$
and exchanging $\vec{E}_{1}$ and $\vec{E}_{2}$. Here $\vec{d}_{m,n}\equiv\langle m|\vec{d}|n\rangle$
is the electric dipole transition matrix element, and $A^{\left(\omega_{3}\right)}$,
$A_{n}^{\left(\omega_{1},\omega_{2}\right)}$, and $A_{n}^{\left(\omega_{2},\omega_{1}\right)}$
are coupling coefficients that depend on detunings and on the envelopes
of the pulses. The orientation averaged population is given by $P_{e}\equiv\int\mathrm{d}\varrho|a_{e}|^{2}$,
which contains contributions from all three sets of pathways and their
interferences. These can be classified according to the number of
photons {[}i.e. electric-dipole products ($\vec{d}\cdot\vec{E})${]}
they contain. If they contain an even number of photons they are not
enantiosensitive (i.e. they are the same in opposite enantiomers)
whereas if they contain an odd number of photons they are enantiosensitive
(i.e. have opposite signs in opposite enantiomers) \citep{ordonezGeometricApproachDecoding2022}.
The latter result from the interference between the one photon pathway
and either of the two-photon pathways. Such interferences contain
expressions such as $(\vec{d}_{e,g}\cdot\vec{E}_{3})^{*}(\vec{d}_{e,n}\cdot\vec{E}_{1})(\vec{d}_{n,g}\cdot\vec{E}_{2})$,
involving three photons. The population in R and S enantiomers can
thus be written as $P_{e}^{\mathrm{R}}=\overline{P}_{e}-\frac{1}{2}\Delta P_{e}$
and $P_{e}^{\mathrm{S}}=\overline{P}_{e}+\frac{1}{2}\Delta P_{e}$,
where $\overline{P}_{e}=(P_{e}^{\mathrm{R}}+P_{e}^{\mathrm{S}})/2$
is the non-enantiosensitive part and $\Delta P_{e}/2=(P_{e}^{\mathrm{S}}-P_{e}^{\mathrm{R}})/2$
is the enantiosensitive part. The latter is given by

\begin{align}
\frac{1}{2}\Delta P_{e} & =2\mathrm{Re}\left\{ \int\mathrm{d}\varrho\,a_{e}^{\left(\omega_{3}\right)*}\sum_{n}\left(a_{e,n}^{\left(\omega_{1},\omega_{2}\right)}+a_{e,n}^{\left(\omega_{2},\omega_{1}\right)}\right)\right\} \\
 & =2\mathrm{Re}\left\{ A^{\left(\omega_{3}\right)*}\sum_{n}\left(A_{n}^{\left(\omega_{1},\omega_{2}\right)}-A_{n}^{\left(\omega_{2},\omega_{1}\right)}\right)\left[\vec{d}_{e,g}^{\,\mathrm{R}}\cdot\left(\vec{d}_{e,n}^{\,\mathrm{R}}\times\vec{d}_{n,g}^{\,\mathrm{R}}\right)\right]\left[\vec{E}_{3}^{*}\cdot\left(\vec{E}_{1}\times\vec{E}_{2}\right)\right]\right\} ,\label{eq:DeltaPe_methods}
\end{align}

where $\vec{d}_{m,n}^{\mathrm{\,R}}$ are the dipoles of the R enantiomer.
The term $[\vec{d}_{e,g}^{\,\mathrm{R}}\cdot(\vec{d}_{e,n}^{\,\mathrm{R}}\times\vec{d}_{n,g}^{\,\mathrm{R}})]=-[\vec{d}_{e,g}^{\,\mathrm{S}}\cdot(\vec{d}_{e,n}^{\,\mathrm{S}}\times\vec{d}_{n,g}^{\,\mathrm{S}})]$
encodes the molecular chirality while $h^{\left(3\right)}\equiv[\vec{E}_{3}^{*}\cdot(\vec{E}_{1}\times\vec{E}_{2})]$
encodes the chirality of the field. The triple products in Eq. (\ref{eq:DeltaPe_methods})
result from molecular orientation averaging \citep{andrewsThreeDimensionalRotational1977}.
Equation (\ref{eq:DeltaPe_methods}) reduces to Eqs. (\ref{eq:DeltaP_3})
and (\ref{eq:h3}) by introducing the shorthand notation

\begin{equation}
g^{\left(3\right)}\equiv4A^{\left(\omega_{3}\right)*}\sum_{n}\left(A_{n}^{\left(\omega_{1},\omega_{2}\right)}-A_{n}^{\left(\omega_{2},\omega_{1}\right)}\right)\left[\vec{d}_{e,g}^{\,\mathrm{R}}\cdot\left(\vec{d}_{e,n}^{\,\mathrm{R}}\times\vec{d}_{n,g}^{\,\mathrm{R}}\right)\right].
\end{equation}

The minus sign in $(A_{n}^{\left(\omega_{1},\omega_{2}\right)}-A_{n}^{\left(\omega_{2},\omega_{1}\right)})$
is related to the opposite photon ordering of both contributions and
the anti-commutativity of the triple product. In general, $A_{n}^{\left(\omega_{1},\omega_{2}\right)}\neq A_{n}^{\left(\omega_{2},\omega_{1}\right)}$,
i.e. the energy structure favours a particular photon ordering. The
difference between $A_{n}^{\left(\omega_{1},\omega_{2}\right)}$ and
$A_{n}^{\left(\omega_{2},\omega_{1}\right)}$ is very pronounced when
the transition to the intermediate state is resonant for one of the
two photon orderings. In the other extreme, if the photon energy difference
is very small with respect to the detunings, i.e. if $|\omega_{1}-\omega_{2}|\ll|\omega_{1}-\omega_{n0}|$,
the two photon orderings become indistinguishable, $A_{n}^{\left(\omega_{1},\omega_{2}\right)}\approx A_{n}^{\left(\omega_{2},\omega_{1}\right)}$,
$g^{\left(3\right)}$ vanishes, and $\Delta P_{e}=0$. 

For illustrative purposes, we include here the coupling coefficients
relevant for identical Gaussian envelopes $\mathcal{E}_{m}(t)=$ $\exp(-t^{2}/2\tau^{2})$ 

\begin{equation}
A^{\left(\omega_{3}\right)}=i\sqrt{\frac{\pi}{2}}\tau e^{-\tau^{2}\Delta_{eg}^{2}/2},
\end{equation}

\begin{equation}
A_{n}^{\left(\omega_{1},\omega_{2}\right)}=\frac{i\tau^{2}}{4}e^{-\tau^{2}\Delta_{eg}^{2}/4}\left[2\sqrt{\pi}D_{+}(\xi)+i\pi e^{-\xi^{2}}\right],
\end{equation}

where $\Delta_{eg}\equiv\omega_{eg}-\omega_{3}$, $\xi\equiv(2\Delta_{n}-\Delta_{eg})\tau/2$,
$\Delta_{n}\equiv\omega_{n0}-\omega_{2}$, and $D_{+}(x)\equiv e^{-x^{2}}\int_{0}^{x}e^{t^{2}}\mathrm{d}t$
is the Dawson function. The expression for $A_{n}^{\left(\omega_{2},\omega_{1}\right)}$
can be obtained from $A_{n}^{\left(\omega_{1},\omega_{2}\right)}$
by exchanging $\omega_{1}$ and $\omega_{2}$.

\subsubsection*{Analytical model for 2- vs 3-photon isotropic chiral coherent control}

In this case, the frequencies are such that $\omega_{3}=2\omega_{2}$
and $\omega_{1}+2\omega_{2}$ is resonant with the transition from
the ground state $|g\rangle$ to the excited state $|e\rangle$. This
leads to six sets of excitation pathways: two 2-photon pathways, $|g\rangle\overset{\omega_{3}}{\rightarrow}|m\rangle\overset{\omega_{1}}{\rightarrow}|e\rangle$
and $|g\rangle\overset{\omega_{1}}{\rightarrow}|m\rangle\overset{\omega_{3}}{\rightarrow}|e\rangle$,
and three 3-photon pathways $|g\rangle\overset{\omega_{2}}{\rightarrow}|n\rangle\overset{\omega_{2}}{\rightarrow}|l\rangle\overset{\omega_{1}}{\rightarrow}|e\rangle$,
$|g\rangle\overset{\omega_{2}}{\rightarrow}|n\rangle\overset{\omega_{1}}{\rightarrow}|l\rangle\overset{\omega_{2}}{\rightarrow}|e\rangle$,
and $|g\rangle\overset{\omega_{1}}{\rightarrow}|n\rangle\overset{\omega_{2}}{\rightarrow}|l\rangle\overset{\omega_{2}}{\rightarrow}|e\rangle$,
where $|m\rangle$, $|n\rangle$, and $|l\rangle$ are intermediate
states that have to be summed over. The probability amplitude for
state $|e\rangle$ after the interaction is given by

\begin{equation}
a_{e}=\sum_{m}a_{e,m}+\sum_{nl}a_{e,nl},
\end{equation}

where perturbation theory yields 

\begin{equation}
a_{e,m}=\sum_{\pi}A_{m}^{\pi\left(\omega_{1},\omega_{3}\right)}\left(\vec{d}_{e,m}\cdot\vec{E}_{1}^{\pi}\right)\left(\vec{d}_{m,g}\cdot\vec{E}_{3}^{\pi}\right),
\end{equation}

\begin{equation}
a_{e,nl}=\sum_{\pi^{\prime}}A_{nl}^{\pi^{\prime}\left(\omega_{1},\omega_{2},\omega_{2}\right)}\left(\vec{d}_{e,n}\cdot\vec{E}_{1}^{\pi^{\prime}}\right)\left(\vec{d}_{n,l}\cdot\vec{E}_{2}^{\pi^{\prime}}\right)\left(\vec{d}_{l,g}\cdot\vec{E}_{2}^{\pi^{\prime}}\right),
\end{equation}

and the sums are over permutations $\pi$ and $\pi^{\prime}$ of the
photon orderings in each pathway. The interference between 2- and
a 3-photon pathways will lead to an enantioselective contribution
to the orientation-averaged population of the form

\begin{align}
\frac{1}{2}\Delta P_{e} & =2\mathrm{Re}\bigg\{\sum_{m,n,l}\sum_{\pi,\pi^{\prime}}A_{m}^{\pi\left(\omega_{1},\omega_{3}\right)*}A_{nl}^{\pi^{\prime}\left(\omega_{1},\omega_{2},\omega_{2}\right)}\int\mathrm{d}\varrho\left(\vec{d}_{e,m}\cdot\vec{E}_{1}^{\pi}\right)^{*}\left(\vec{d}_{m,g}\cdot\vec{E}_{3}^{\pi}\right)^{*}\left(\vec{d}_{e,n}\cdot\vec{E}_{1}^{\pi^{\prime}}\right)\left(\vec{d}_{n,l}\cdot\vec{E}_{2}^{\pi^{\prime}}\right)\left(\vec{d}_{l,g}\cdot\vec{E}_{2}^{\pi^{\prime}}\right)\bigg\}.\label{eq:DeltaP_2vs3_unsolved}
\end{align}
The integral over orientations can be performed analytically following
Ref. \citep{andrewsThreeDimensionalRotational1977} (see e.g. the
SI of \citep{ordonezGeometricApproachDecoding2022} for a worked out
example). Each integral over orientations yields a result of the form
$\vec{F}\cdot M\vec{G}$, where $\vec{F}$ ($\vec{G}$) is a six-dimensional
vector of rotational invariants involving triple products and dot
products between the electric fields (electric dipole) vectors only,
and $M$ is a matrix coupling the molecular and the field invariants.
See e.g. Eqs. (\ref{eq:F})-(\ref{eq:G}). The different permutations
in Eq. (\ref{eq:DeltaP_2vs3_unsolved}) will yield different $\vec{F}$'s,
which complicates interpretation. However, since a permutation of
the electric fields $\vec{E}_{j}$ in the integral over orientations
is equivalent to a corresponding permutation of the electric dipoles,
we choose to keep the $\vec{E}_{j}$'s fixed and permute the dipoles
instead. That is, we consider integrals with the form 

\begin{equation}
\int\mathrm{d}\varrho\left(\vec{d}_{e,m}^{\pi}\cdot\vec{E}_{1}\right)^{*}\left(\vec{d}_{m,g}^{\pi}\cdot\vec{E}_{3}\right)^{*}\left(\vec{d}_{e,n}^{\pi^{\prime}}\cdot\vec{E}_{1}\right)\left(\vec{d}_{n,l}^{\pi^{\prime}}\cdot\vec{E}_{2}\right)\left(\vec{d}_{l,g}^{\pi^{\prime}}\cdot\vec{E}_{2}\right),
\end{equation}

where the permutations are now among the molecular dipoles. To simplify
the expression for $\vec{F}$, we reorder the dot products in the
integral and apply Eq. (24) in Ref. \citep{andrewsThreeDimensionalRotational1977}
according to 

\begin{equation}
\int\mathrm{d}\varrho\left(\vec{d}_{n,l}^{\pi^{\prime}}\cdot\vec{E}_{2}\right)\left(\vec{d}_{l,g}^{\pi^{\prime}}\cdot\vec{E}_{2}\right)\left(\vec{d}_{e,n}^{\pi^{\prime}}\cdot\vec{E}_{1}\right)\left(\vec{d}_{e,m}^{\pi}\cdot\vec{E}_{1}\right)^{*}\left(\vec{d}_{m,g}^{\pi}\cdot\vec{E}_{3}\right)^{*}=\vec{F}\cdot M\vec{G}_{\pi,\pi^{\prime}},
\end{equation}

where

\begin{equation}
\vec{F}=\left(\begin{array}{c}
0\\
0\\
0\\
\five{\vec{E}_{2}}{\vec{E}_{1}}{\vec{E}_{1}^{*}}{\vec{E}_{2}}{\vec{E}_{3}^{*}}\\
\five{\vec{E}_{2}}{\vec{E}_{1}}{\vec{E}_{3}^{*}}{\vec{E}_{2}}{\vec{E}_{1}^{*}}\\
\five{\vec{E}_{2}}{\vec{E}_{1}^{*}}{\vec{E}_{3}^{*}}{\vec{E}_{2}}{\vec{E}_{1}}
\end{array}\right),\label{eq:F}
\end{equation}

\begin{equation}
M=\frac{1}{30}\left(\begin{array}{cccccc}
3 & -1 & -1 & 1 & 1 & 0\\
-1 & 3 & -1 & -1 & 0 & 1\\
-1 & -1 & 3 & 0 & -1 & -1\\
1 & -1 & 0 & 3 & -1 & 1\\
1 & 0 & -1 & -1 & 3 & -1\\
0 & 1 & -1 & 1 & -1 & 3
\end{array}\right),
\end{equation}

\begin{equation}
\vec{G}_{\pi,\pi^{\prime}}=\Five{\vec{d}_{n,l}^{\pi^{\prime}}}{\vec{d}_{l,g}^{\pi^{\prime}}}{\vec{d}_{e,n}^{\pi^{\prime}}}{\vec{d}_{e,m}^{\pi}}{\vec{d}_{m,g}^{\pi}}.\label{eq:G}
\end{equation}

Replacing in the expression for the enantiosensitive contribution
to the population, we obtain 

\begin{align}
\frac{1}{2}\Delta P_{e} & =2\mathrm{Re}\bigg\{\vec{F}\cdot M\sum_{m,n,p}\sum_{\pi,\pi^{\prime}}A_{m}^{\pi\left(\omega_{1},\omega_{3}\right)*}A_{n,l}^{\pi^{\prime}\left(\omega_{1},\omega_{2},\omega_{2}\right)}\vec{G}_{\pi,\pi^{\prime}}\bigg\}.
\end{align}

This equation reduces to Eqs. (\ref{eq:DeltaP_5}) and (\ref{eq:h5})
if we define

\begin{equation}
\vec{g}^{\left(5\right)}\equiv4M^{\prime}\sum_{m,n,p}\sum_{\pi,\pi^{\prime}}A_{m}^{\pi\left(\omega_{1},\omega_{3}\right)*}A_{n,l}^{\pi^{\prime}\left(\omega_{1},\omega_{2},\omega_{2}\right)}\vec{G}_{\pi,\pi^{\prime}},
\end{equation}
 where $\vec{h}^{\left(5\right)}$ is given by the non-zero components
of $\vec{F}$ and $M^{\prime}$ is the matrix consisting of the 4th
to 6th rows of $M$.

\section*{Acknowledgements}

A. O. acknowledges funding from the European Union's Horizon 2020
research and innovation programme under the Marie Sk\l odowska-Curie
grant agreement No 101029393, and from UK Research and Innovation
Engineering and Physical Sciences Research Council EP/Z002834/1. P.
V. Z. is grateful to the National Science Foundation for its partial
support through the Grants CHE-2054616 and CHE-2054604 and the Simons
Foundation for the computational resources. A. O. and D. A. acknowledge
funding from the Royal Society URF$\backslash$R1$\backslash$201333,
URF$\backslash$ERE$\backslash$210358 and URF$\backslash$ERE$\backslash$231177.

\bibliographystyle{naturemag}
\bibliography{/Users/andres/Documents/Books/MyLibrary}

\begin{thebibliography}{10}
\expandafter\ifx\csname url\endcsname\relax
  \def\url#1{\texttt{#1}}\fi
\expandafter\ifx\csname urlprefix\endcsname\relax\def\urlprefix{URL }\fi
\providecommand{\bibinfo}[2]{#2}
\providecommand{\eprint}[2][]{\url{#2}}

\bibitem{hellNobelLectureNanoscopy2015}
\bibinfo{author}{Hell, S.~W.}
\newblock \bibinfo{title}{Nobel {{Lecture}}: {{Nanoscopy}} with freely
  propagating light}.
\newblock \emph{\bibinfo{journal}{Reviews of Modern Physics}}
  \textbf{\bibinfo{volume}{87}}, \bibinfo{pages}{1169--1181}
  (\bibinfo{year}{2015}).

\bibitem{betzigNobelLectureSingle2015}
\bibinfo{author}{Betzig, E.}
\newblock \bibinfo{title}{Nobel {{Lecture}}: {{Single}} molecules, cells, and
  super-resolution optics}.
\newblock \emph{\bibinfo{journal}{Reviews of Modern Physics}}
  \textbf{\bibinfo{volume}{87}}, \bibinfo{pages}{1153--1168}
  (\bibinfo{year}{2015}).

\bibitem{moernerNobelLectureSinglemolecule2015}
\bibinfo{author}{Moerner, W. E. W.~E.}
\newblock \bibinfo{title}{Nobel {{Lecture}}: {{Single-molecule}} spectroscopy,
  imaging, and photocontrol: {{Foundations}} for super-resolution microscopy}.
\newblock \emph{\bibinfo{journal}{Reviews of Modern Physics}}
  \textbf{\bibinfo{volume}{87}}, \bibinfo{pages}{1183--1212}
  (\bibinfo{year}{2015}).

\bibitem{lauterburAllScienceInterdisciplinary2005}
\bibinfo{author}{Lauterbur, P.~C.}
\newblock \bibinfo{title}{All {{Science Is Interdisciplinary}}---{{From
  Magnetic Moments}} to {{Molecules}} to {{Men}} ({{Nobel Lecture}})}.
\newblock \emph{\bibinfo{journal}{Angewandte Chemie International Edition}}
  \textbf{\bibinfo{volume}{44}}, \bibinfo{pages}{1004--1011}
  (\bibinfo{year}{2005}).

\bibitem{mansfieldSnapshotMagneticResonance2004}
\bibinfo{author}{Mansfield, P.}
\newblock \bibinfo{title}{Snapshot {{Magnetic Resonance Imaging}} ({{Nobel
  Lecture}})}.
\newblock \emph{\bibinfo{journal}{Angewandte Chemie International Edition}}
  \textbf{\bibinfo{volume}{43}}, \bibinfo{pages}{5456--5464}
  (\bibinfo{year}{2004}).

\bibitem{liuDetectionAnalysisChiral2023}
\bibinfo{author}{Liu, Y.}, \bibinfo{author}{Wu, Z.},
  \bibinfo{author}{Armstrong, D.~W.}, \bibinfo{author}{Wolosker, H.} \&
  \bibinfo{author}{Zheng, Y.}
\newblock \bibinfo{title}{Detection and analysis of chiral molecules as disease
  biomarkers}.
\newblock \emph{\bibinfo{journal}{Nature Reviews Chemistry}}
  \textbf{\bibinfo{volume}{7}}, \bibinfo{pages}{355--373}
  (\bibinfo{year}{2023}).

\bibitem{brandtAddedValueSmallmolecule2017}
\bibinfo{author}{Brandt, J.~R.}, \bibinfo{author}{Salerno, F.} \&
  \bibinfo{author}{Fuchter, M.~J.}
\newblock \bibinfo{title}{The added value of small-molecule chirality in
  technological applications}.
\newblock \emph{\bibinfo{journal}{Nature Reviews Chemistry}}
  \textbf{\bibinfo{volume}{1}}, \bibinfo{pages}{1--12} (\bibinfo{year}{2017}).

\bibitem{feringaArtBuildingSmall2017}
\bibinfo{author}{Feringa, B.~L.}
\newblock \bibinfo{title}{The {{Art}} of {{Building Small}}: {{From Molecular
  Switches}} to {{Motors}} ({{Nobel Lecture}})}.
\newblock \emph{\bibinfo{journal}{Angewandte Chemie International Edition}}
  \textbf{\bibinfo{volume}{56}}, \bibinfo{pages}{11060--11078}
  (\bibinfo{year}{2017}).

\bibitem{feringaMolecularSwitches2001}
\bibinfo{author}{Feringa, B.~L.}
\newblock \emph{\bibinfo{title}{Molecular {{Switches}}}}
  (\bibinfo{publisher}{Wiley-VCH Verlag GmbH}, \bibinfo{year}{2001}).

\bibitem{naamanChiralMoleculesElectron2019}
\bibinfo{author}{Naaman, R.}, \bibinfo{author}{Paltiel, Y.} \&
  \bibinfo{author}{Waldeck, D.~H.}
\newblock \bibinfo{title}{Chiral molecules and the electron spin}.
\newblock \emph{\bibinfo{journal}{Nature Reviews Chemistry}}
  \textbf{\bibinfo{volume}{3}}, \bibinfo{pages}{250--260}
  (\bibinfo{year}{2019}).

\bibitem{choBioinspiredChiralInorganic2023}
\bibinfo{author}{Cho, N.~H.} \emph{et~al.}
\newblock \bibinfo{title}{Bioinspired chiral inorganic nanomaterials}.
\newblock \emph{\bibinfo{journal}{Nature Reviews Bioengineering}}
  \textbf{\bibinfo{volume}{1}}, \bibinfo{pages}{88} (\bibinfo{year}{2023}).

\bibitem{ayusoUltrafastChiralityRoad2022a}
\bibinfo{author}{Ayuso, D.}, \bibinfo{author}{Ordonez, A.~F.} \&
  \bibinfo{author}{Smirnova, O.}
\newblock \bibinfo{title}{Ultrafast chirality: The road to efficient chiral
  measurements}.
\newblock \emph{\bibinfo{journal}{Physical Chemistry Chemical Physics}}
  \textbf{\bibinfo{volume}{24}}, \bibinfo{pages}{26962--26991}
  (\bibinfo{year}{2022}).

\bibitem{rouxelMolecularChiralityIts2022}
\bibinfo{author}{Rouxel, J.~R.} \& \bibinfo{author}{Mukamel, S.}
\newblock \bibinfo{title}{Molecular {{Chirality}} and {{Its Monitoring}} by
  {{Ultrafast X-ray Pulses}}}.
\newblock \emph{\bibinfo{journal}{Chemical Reviews}}  (\bibinfo{year}{2022}).

\bibitem{beginNonlinearHelicalDichroism2023}
\bibinfo{author}{B{\'e}gin, J.-L.} \emph{et~al.}
\newblock \bibinfo{title}{Nonlinear helical dichroism in chiral and achiral
  molecules}.
\newblock \emph{\bibinfo{journal}{Nature Photonics}}
  \textbf{\bibinfo{volume}{17}}, \bibinfo{pages}{82--88}
  (\bibinfo{year}{2023}).

\bibitem{sparlingTwoDecadesImaging2025}
\bibinfo{author}{Sparling, C.} \& \bibinfo{author}{Townsend, D.}
\newblock \bibinfo{title}{Two decades of imaging photoelectron circular
  dichroism: From first principles to future perspectives}.
\newblock \emph{\bibinfo{journal}{Physical Chemistry Chemical Physics}}
  (\bibinfo{year}{2025}).

\bibitem{koumarianouAssignmentfreeChiralityDetection2022}
\bibinfo{author}{Koumarianou, G.}, \bibinfo{author}{Wang, I.},
  \bibinfo{author}{Satterthwaite, L.} \& \bibinfo{author}{Patterson, D.}
\newblock \bibinfo{title}{Assignment-free chirality detection in unknown
  samples via microwave three-wave mixing}.
\newblock \emph{\bibinfo{journal}{Communications Chemistry}}
  \textbf{\bibinfo{volume}{5}}, \bibinfo{pages}{1--7} (\bibinfo{year}{2022}).

\bibitem{tutunnikovObservationPersistentOrientation2020}
\bibinfo{author}{Tutunnikov, I.} \emph{et~al.}
\newblock \bibinfo{title}{Observation of persistent orientation of chiral
  molecules by a laser field with twisted polarization}.
\newblock \emph{\bibinfo{journal}{Physical Review A}}
  \textbf{\bibinfo{volume}{101}}, \bibinfo{pages}{021403}
  (\bibinfo{year}{2020}).

\bibitem{schwennickeEnantioselectiveTopologicalFrequency2022}
\bibinfo{author}{Schwennicke, K.} \& \bibinfo{author}{{Yuen-Zhou}, J.}
\newblock \bibinfo{title}{Enantioselective {{Topological Frequency
  Conversion}}}.
\newblock \emph{\bibinfo{journal}{The Journal of Physical Chemistry Letters}}
  \textbf{\bibinfo{volume}{13}}, \bibinfo{pages}{2434--2441}
  (\bibinfo{year}{2022}).

\bibitem{l.greenfieldPathwaysIncreaseDissymmetry2021}
\bibinfo{author}{L.~Greenfield, J.} \emph{et~al.}
\newblock \bibinfo{title}{Pathways to increase the dissymmetry in the
  interaction of chiral light and chiral molecules}.
\newblock \emph{\bibinfo{journal}{Chemical Science}}
  \textbf{\bibinfo{volume}{12}}, \bibinfo{pages}{8589--8602}
  (\bibinfo{year}{2021}).

\bibitem{raucciChiralPhotochemistryAchiral2022}
\bibinfo{author}{Raucci, U.}, \bibinfo{author}{Weir, H.},
  \bibinfo{author}{Bannwarth, C.}, \bibinfo{author}{Sanchez, D.~M.} \&
  \bibinfo{author}{Mart{\'i}nez, T.~J.}
\newblock \bibinfo{title}{Chiral photochemistry of achiral molecules}.
\newblock \emph{\bibinfo{journal}{Nature Communications}}
  \textbf{\bibinfo{volume}{13}}, \bibinfo{pages}{2091} (\bibinfo{year}{2022}).

\bibitem{barronMolecularLightScattering2004}
\bibinfo{author}{Barron, L.~D.}
\newblock \emph{\bibinfo{title}{Molecular Light Scattering and Optical
  Activity}} (\bibinfo{publisher}{Cambridge University Press},
  \bibinfo{year}{2004}), \bibinfo{edition}{2nd} edn.

\bibitem{berovaComprehensiveChiropticalSpectroscopy2012}
\bibinfo{author}{Berova, N.}, \bibinfo{author}{Polavarapu, P.~L.},
  \bibinfo{author}{Nakanishi, K.} \& \bibinfo{author}{Woody, R.~W.}
\newblock \emph{\bibinfo{title}{Comprehensive {{Chiroptical Spectroscopy}}}}
  (\bibinfo{publisher}{Wiley}, \bibinfo{address}{Hoboken, New Jersey},
  \bibinfo{year}{2012}).

\bibitem{tangEnhancedEnantioselectivityExcitation2011}
\bibinfo{author}{Tang, Y.} \& \bibinfo{author}{Cohen, A.~E.}
\newblock \bibinfo{title}{Enhanced {{Enantioselectivity}} in {{Excitation}} of
  {{Chiral Molecules}} by {{Superchiral Light}}}.
\newblock \emph{\bibinfo{journal}{Science}} \textbf{\bibinfo{volume}{332}},
  \bibinfo{pages}{333--336} (\bibinfo{year}{2011}).

\bibitem{heDissymmetryEnhancementEnantioselective2018}
\bibinfo{author}{He, C.} \emph{et~al.}
\newblock \bibinfo{title}{Dissymmetry enhancement in enantioselective synthesis
  of helical polydiacetylene by application of superchiral light}.
\newblock \emph{\bibinfo{journal}{Nature Communications}}
  \textbf{\bibinfo{volume}{9}}, \bibinfo{pages}{5117} (\bibinfo{year}{2018}).

\bibitem{shapiroQuantumControlMolecular2012}
\bibinfo{author}{Shapiro, M.} \& \bibinfo{author}{Brumer, P.}
\newblock \emph{\bibinfo{title}{Quantum {{Control}} of {{Molecular
  Processes}}}} (\bibinfo{publisher}{Wiley-VCH}, \bibinfo{address}{Weinheim},
  \bibinfo{year}{2012}), \bibinfo{edition}{2nd} edn.

\bibitem{gerbasiTheoryTwoStep2004}
\bibinfo{author}{Gerbasi, D.}, \bibinfo{author}{Brumer, P.},
  \bibinfo{author}{Thanopulos, I.}, \bibinfo{author}{Kr{\'a}l, P.} \&
  \bibinfo{author}{Shapiro, M.}
\newblock \bibinfo{title}{Theory of the two step enantiomeric purification of
  1,3 dimethylallene}.
\newblock \emph{\bibinfo{journal}{The Journal of Chemical Physics}}
  \textbf{\bibinfo{volume}{120}}, \bibinfo{pages}{11557--11563}
  (\bibinfo{year}{2004}).

\bibitem{leibscherFullQuantumControl2022}
\bibinfo{author}{Leibscher, M.} \emph{et~al.}
\newblock \bibinfo{title}{Full quantum control of enantiomer-selective state
  transfer in chiral molecules despite degeneracy}.
\newblock \emph{\bibinfo{journal}{Communications Physics}}
  \textbf{\bibinfo{volume}{5}}, \bibinfo{pages}{1--16} (\bibinfo{year}{2022}).

\bibitem{eibenbergerEnantiomerSpecificStateTransfer2017}
\bibinfo{author}{Eibenberger, S.}, \bibinfo{author}{Doyle, J.} \&
  \bibinfo{author}{Patterson, D.}
\newblock \bibinfo{title}{Enantiomer-{{Specific State Transfer}} of {{Chiral
  Molecules}}}.
\newblock \emph{\bibinfo{journal}{Physical Review Letters}}
  \textbf{\bibinfo{volume}{118}}, \bibinfo{pages}{123002}
  (\bibinfo{year}{2017}).

\bibitem{perezCoherentEnantiomerSelectivePopulation2017}
\bibinfo{author}{P{\'e}rez, C.} \emph{et~al.}
\newblock \bibinfo{title}{Coherent {{Enantiomer-Selective Population Enrichment
  Using Tailored Microwave Fields}}}.
\newblock \emph{\bibinfo{journal}{Angewandte Chemie International Edition}}
  \textbf{\bibinfo{volume}{56}}, \bibinfo{pages}{12512--12517}
  (\bibinfo{year}{2017}).

\bibitem{perezStateSpecificEnrichmentChiral2018}
\bibinfo{author}{P{\'e}rez, C.}, \bibinfo{author}{Steber, A.~L.},
  \bibinfo{author}{Krin, A.} \& \bibinfo{author}{Schnell, M.}
\newblock \bibinfo{title}{State-{{Specific Enrichment}} of {{Chiral
  Conformers}} with {{Microwave Spectroscopy}}}.
\newblock \emph{\bibinfo{journal}{The Journal of Physical Chemistry Letters}}
  \textbf{\bibinfo{volume}{9}}, \bibinfo{pages}{4539--4543}
  (\bibinfo{year}{2018}).

\bibitem{sunInducingTransientEnantiomeric2023}
\bibinfo{author}{Sun, W.} \emph{et~al.}
\newblock \bibinfo{title}{Inducing transient enantiomeric excess in a molecular
  quantum racemic mixture with microwave fields}.
\newblock \emph{\bibinfo{journal}{Nature Communications}}
  \textbf{\bibinfo{volume}{14}}, \bibinfo{pages}{934} (\bibinfo{year}{2023}).

\bibitem{leeQuantitativeStudyEnantiomerSpecific2022}
\bibinfo{author}{Lee, J.} \emph{et~al.}
\newblock \bibinfo{title}{Quantitative {{Study}} of {{Enantiomer-Specific State
  Transfer}}}.
\newblock \emph{\bibinfo{journal}{Physical Review Letters}}
  \textbf{\bibinfo{volume}{128}}, \bibinfo{pages}{173001}
  (\bibinfo{year}{2022}).

\bibitem{leeNearcompleteChiralSelection2024}
\bibinfo{author}{Lee, J.}, \bibinfo{author}{Abdiha, E.},
  \bibinfo{author}{Sartakov, B.~G.}, \bibinfo{author}{Meijer, G.} \&
  \bibinfo{author}{{Eibenberger-Arias}, S.}
\newblock \bibinfo{title}{Near-complete chiral selection in rotational quantum
  states}.
\newblock \emph{\bibinfo{journal}{Nature Communications}}
  \textbf{\bibinfo{volume}{15}}, \bibinfo{pages}{7441} (\bibinfo{year}{2024}).

\bibitem{schuurmanDynamicsConicalIntersections2018}
\bibinfo{author}{Schuurman, M.~S.} \& \bibinfo{author}{Stolow, A.}
\newblock \bibinfo{title}{Dynamics at {{Conical Intersections}}}.
\newblock \emph{\bibinfo{journal}{Annual Review of Physical Chemistry}}
  \textbf{\bibinfo{volume}{69}}, \bibinfo{pages}{427--450}
  (\bibinfo{year}{2018}).

\bibitem{borneUltrafastElectronicRelaxation2024}
\bibinfo{author}{Borne, K.~D.} \emph{et~al.}
\newblock \bibinfo{title}{Ultrafast electronic relaxation pathways of the
  molecular photoswitch quadricyclane}.
\newblock \emph{\bibinfo{journal}{Nature Chemistry}}
  \textbf{\bibinfo{volume}{16}}, \bibinfo{pages}{499--505}
  (\bibinfo{year}{2024}).

\bibitem{brimioulleEnantioselectiveCatalysisPhotochemical2015}
\bibinfo{author}{Brimioulle, R.}, \bibinfo{author}{Lenhart, D.},
  \bibinfo{author}{Maturi, M.~M.} \& \bibinfo{author}{Bach, T.}
\newblock \bibinfo{title}{Enantioselective {{Catalysis}} of {{Photochemical
  Reactions}}}.
\newblock \emph{\bibinfo{journal}{Angewandte Chemie International Edition}}
  \textbf{\bibinfo{volume}{54}}, \bibinfo{pages}{3872--3890}
  (\bibinfo{year}{2015}).

\bibitem{tarafderChiralChromatographyMethod2021}
\bibinfo{author}{Tarafder, A.} \& \bibinfo{author}{Miller, L.}
\newblock \bibinfo{title}{Chiral chromatography method screening strategies:
  {{Past}}, present and future}.
\newblock \emph{\bibinfo{journal}{Journal of Chromatography A}}
  \textbf{\bibinfo{volume}{1638}}, \bibinfo{pages}{461878}
  (\bibinfo{year}{2021}).

\bibitem{genzinkChiralPhotocatalystStructures2022}
\bibinfo{author}{Genzink, M.~J.}, \bibinfo{author}{Kidd, J.~B.},
  \bibinfo{author}{Swords, W.~B.} \& \bibinfo{author}{Yoon, T.~P.}
\newblock \bibinfo{title}{Chiral {{Photocatalyst Structures}} in {{Asymmetric
  Photochemical Synthesis}}}.
\newblock \emph{\bibinfo{journal}{Chemical Reviews}}
  \textbf{\bibinfo{volume}{122}}, \bibinfo{pages}{1654--1716}
  (\bibinfo{year}{2022}).

\bibitem{fattahiThirdgenerationFemtosecondTechnology2014}
\bibinfo{author}{Fattahi, H.} \emph{et~al.}
\newblock \bibinfo{title}{Third-generation femtosecond technology}.
\newblock \emph{\bibinfo{journal}{Optica}} \textbf{\bibinfo{volume}{1}},
  \bibinfo{pages}{45--63} (\bibinfo{year}{2014}).

\bibitem{burgerCompactFlexibleHarmonic2017}
\bibinfo{author}{Burger, C.} \emph{et~al.}
\newblock \bibinfo{title}{Compact and flexible harmonic generator and
  three-color synthesizer for femtosecond coherent control and time-resolved
  studies}.
\newblock \emph{\bibinfo{journal}{Optics Express}}
  \textbf{\bibinfo{volume}{25}}, \bibinfo{pages}{31130--31139}
  (\bibinfo{year}{2017}).

\bibitem{brixnerPhotoselectiveAdaptiveFemtosecond2001}
\bibinfo{author}{Brixner, T.}, \bibinfo{author}{Damrauer, N.~H.},
  \bibinfo{author}{Niklaus, P.} \& \bibinfo{author}{Gerber, G.}
\newblock \bibinfo{title}{Photoselective adaptive femtosecond quantum control
  in the liquid phase}.
\newblock \emph{\bibinfo{journal}{Nature}} \textbf{\bibinfo{volume}{414}},
  \bibinfo{pages}{57--60} (\bibinfo{year}{2001}).

\bibitem{hokiSelectivePreparationEnantiomers2001}
\bibinfo{author}{Hoki, K.}, \bibinfo{author}{Kr{\"o}ner, D.} \&
  \bibinfo{author}{Manz, J.}
\newblock \bibinfo{title}{Selective preparation of enantiomers from a racemate
  by laser pulses: Model simulation for oriented atropisomers with coupled
  rotations and torsions}.
\newblock \emph{\bibinfo{journal}{Chemical Physics}}
  \textbf{\bibinfo{volume}{267}}, \bibinfo{pages}{59--79}
  (\bibinfo{year}{2001}).

\bibitem{yachmenevFieldInducedDiastereomersChiral2019}
\bibinfo{author}{Yachmenev, A.}, \bibinfo{author}{Onvlee, J.},
  \bibinfo{author}{Zak, E.}, \bibinfo{author}{Owens, A.} \&
  \bibinfo{author}{K{\"u}pper, J.}
\newblock \bibinfo{title}{Field-{{Induced Diastereomers}} for {{Chiral
  Separation}}}.
\newblock \emph{\bibinfo{journal}{Physical Review Letters}}
  \textbf{\bibinfo{volume}{123}}, \bibinfo{pages}{243202}
  (\bibinfo{year}{2019}).

\bibitem{vitanovHighlyEfficientDetection2019}
\bibinfo{author}{Vitanov, N.~V.} \& \bibinfo{author}{Drewsen, M.}
\newblock \bibinfo{title}{Highly {{Efficient Detection}} and {{Separation}} of
  {{Chiral Molecules}} through {{Shortcuts}} to {{Adiabaticity}}}.
\newblock \emph{\bibinfo{journal}{Physical Review Letters}}
  \textbf{\bibinfo{volume}{122}}, \bibinfo{pages}{173202}
  (\bibinfo{year}{2019}).

\bibitem{torosovEfficientRobustChiral2020}
\bibinfo{author}{Torosov, B.~T.}, \bibinfo{author}{Drewsen, M.} \&
  \bibinfo{author}{Vitanov, N.~V.}
\newblock \bibinfo{title}{Efficient and robust chiral resolution by composite
  pulses}.
\newblock \emph{\bibinfo{journal}{Physical Review A}}
  \textbf{\bibinfo{volume}{101}}, \bibinfo{pages}{063401}
  (\bibinfo{year}{2020}).

\bibitem{neufeldStrongChiralDichroism2021}
\bibinfo{author}{Neufeld, O.}, \bibinfo{author}{H{\"u}bener, H.},
  \bibinfo{author}{Rubio, A.} \& \bibinfo{author}{De~Giovannini, U.}
\newblock \bibinfo{title}{Strong chiral dichroism and enantiopurification in
  above-threshold ionization with locally chiral light}.
\newblock \emph{\bibinfo{journal}{Physical Review Research}}
  \textbf{\bibinfo{volume}{3}}, \bibinfo{pages}{L032006}
  (\bibinfo{year}{2021}).

\bibitem{yePhasematchedLocallyChiral2023}
\bibinfo{author}{Ye, C.}, \bibinfo{author}{Sun, Y.}, \bibinfo{author}{Fu, L.}
  \& \bibinfo{author}{Zhang, X.}
\newblock \bibinfo{title}{Phase-matched locally chiral light for global control
  of chiral light--matter interaction}.
\newblock \emph{\bibinfo{journal}{Optics Letters}}
  \textbf{\bibinfo{volume}{48}}, \bibinfo{pages}{5511--5514}
  (\bibinfo{year}{2023}).

\bibitem{ciamicianChemischeLichtwirkungen1908}
\bibinfo{author}{Ciamician, G.} \& \bibinfo{author}{Silber, P.}
\newblock \bibinfo{title}{Chemische {{Lichtwirkungen}}}.
\newblock \emph{\bibinfo{journal}{Berichte der deutschen chemischen
  Gesellschaft}} \textbf{\bibinfo{volume}{41}}, \bibinfo{pages}{1928--1935}
  (\bibinfo{year}{1908}).

\bibitem{schonbergPreparativeOrganicPhotochemistry1968}
\bibinfo{author}{Sch{\"o}nberg, A.}
\newblock \emph{\bibinfo{title}{Preparative {{Organic Photochemistry}}}}
  (\bibinfo{publisher}{Springer-Verlag New York Inc.}, \bibinfo{year}{1968}).

\bibitem{crimminsEnoneOlefinPhotochemical2004}
\bibinfo{author}{Crimmins, M.~T.} \& \bibinfo{author}{Reinhold, T.~L.}
\newblock \bibinfo{title}{Enone {{Olefin}} [2 + 2] {{Photochemical
  Cycloadditions}}}.
\newblock In \emph{\bibinfo{booktitle}{Organic {{Reactions}}}},
  chap.~\bibinfo{chapter}{2}, \bibinfo{pages}{297--588}
  (\bibinfo{publisher}{John Wiley \& Sons, Ltd}, \bibinfo{year}{2004}).

\bibitem{malatestaLaserinducedCycloadditionsCarvone1982}
\bibinfo{author}{Malatesta, V.}, \bibinfo{author}{Willis, C.} \&
  \bibinfo{author}{Hackett, P.~A.}
\newblock \bibinfo{title}{Laser-induced cycloadditions: The carvone
  photoisomerization}.
\newblock \emph{\bibinfo{journal}{The Journal of Organic Chemistry}}
  \textbf{\bibinfo{volume}{47}}, \bibinfo{pages}{3117--3121}
  (\bibinfo{year}{1982}).

\bibitem{tsipiIntramolecularPhotocycloadditionCarvone1987}
\bibinfo{author}{Tsipi, D.}, \bibinfo{author}{Gegiou, D.} \&
  \bibinfo{author}{Hadjoudis, E.}
\newblock \bibinfo{title}{The intramolecular photocycloaddition carvone
  {$\rightarrow$} carvonecamphor}.
\newblock \emph{\bibinfo{journal}{Journal of Photochemistry}}
  \textbf{\bibinfo{volume}{37}}, \bibinfo{pages}{159--166}
  (\bibinfo{year}{1987}).

\bibitem{zandomeneghiLaserPhotochemistryIntramolecular1980}
\bibinfo{author}{Zandomeneghi, M.}, \bibinfo{author}{Cavazza, M.},
  \bibinfo{author}{Moi, L.} \& \bibinfo{author}{Pietra, F.}
\newblock \bibinfo{title}{Laser photochemistry: {{The}} intramolecular
  cyclization of carvone to carvonecamphor}.
\newblock \emph{\bibinfo{journal}{Tetrahedron Letters}}
  \textbf{\bibinfo{volume}{21}}, \bibinfo{pages}{213--214}
  (\bibinfo{year}{1980}).

\bibitem{buchiPhotochemicalReactionsVII1957}
\bibinfo{author}{B{\"u}chi, G.} \& \bibinfo{author}{Goldman, I.~M.}
\newblock \bibinfo{title}{Photochemical {{Reactions}}. {{VII}}. {{The
  Intramolecular Cyclization}} of {{Carvone}} to {{Carvonecamphor}}}.
\newblock \emph{\bibinfo{journal}{Journal of the American Chemical Society}}
  \textbf{\bibinfo{volume}{79}}, \bibinfo{pages}{4741--4748}
  (\bibinfo{year}{1957}).

\bibitem{brackmannPhotocyclizationCarvoneCarvone1982}
\bibinfo{author}{Brackmann, U.} \& \bibinfo{author}{Sch{\"a}fer, F.~P.}
\newblock \bibinfo{title}{Photocyclization of carvone to carvone camphor using
  rare gas halide lasers}.
\newblock \emph{\bibinfo{journal}{Chemical Physics Letters}}
  \textbf{\bibinfo{volume}{87}}, \bibinfo{pages}{579--581}
  (\bibinfo{year}{1982}).

\bibitem{brimioulleEnantioselectiveLewisAcid2013}
\bibinfo{author}{Brimioulle, R.} \& \bibinfo{author}{Bach, T.}
\newblock \bibinfo{title}{Enantioselective {{Lewis Acid Catalysis}} of
  {{Intramolecular Enone}} [2+2] {{Photocycloaddition Reactions}}}.
\newblock \emph{\bibinfo{journal}{Science}} \textbf{\bibinfo{volume}{342}},
  \bibinfo{pages}{840--843} (\bibinfo{year}{2013}).

\bibitem{sarkarPhotochemicalCycloadditionOrganic2020}
\bibinfo{author}{Sarkar, D.}, \bibinfo{author}{Bera, N.} \&
  \bibinfo{author}{Ghosh, S.}
\newblock \bibinfo{title}{[2+2] {{Photochemical Cycloaddition}} in {{Organic
  Synthesis}}}.
\newblock \emph{\bibinfo{journal}{European Journal of Organic Chemistry}}
  \textbf{\bibinfo{volume}{2020}}, \bibinfo{pages}{1310--1326}
  (\bibinfo{year}{2020}).

\bibitem{wangPhaseControlAbsorption1996}
\bibinfo{author}{Wang, X.}, \bibinfo{author}{Bersohn, R.},
  \bibinfo{author}{Takahashi, K.}, \bibinfo{author}{Kawasaki, M.} \&
  \bibinfo{author}{Kim, H.~L.}
\newblock \bibinfo{title}{Phase control of absorption in large polyatomic
  molecules}.
\newblock \emph{\bibinfo{journal}{The Journal of Chemical Physics}}
  \textbf{\bibinfo{volume}{105}}, \bibinfo{pages}{2992--2997}
  (\bibinfo{year}{1996}).

\bibitem{boydNonlinearOptics2008}
\bibinfo{author}{Boyd, R.}
\newblock \emph{\bibinfo{title}{Nonlinear {{Optics}}}}
  (\bibinfo{publisher}{Elsevier}, \bibinfo{year}{2008}), \bibinfo{edition}{3rd}
  edn.

\bibitem{galinisMicrometerthicknessLiquidSheet2017}
\bibinfo{author}{Galinis, G.} \emph{et~al.}
\newblock \bibinfo{title}{Micrometer-thickness liquid sheet jets flowing in
  vacuum}.
\newblock \emph{\bibinfo{journal}{Review of Scientific Instruments}}
  \textbf{\bibinfo{volume}{88}}, \bibinfo{pages}{083117}
  (\bibinfo{year}{2017}).

\bibitem{luuExtremeUltravioletHigh2018}
\bibinfo{author}{Luu, T.~T.} \emph{et~al.}
\newblock \bibinfo{title}{Extreme--ultraviolet high--harmonic generation in
  liquids}.
\newblock \emph{\bibinfo{journal}{Nature Communications}}
  \textbf{\bibinfo{volume}{9}}, \bibinfo{pages}{3723} (\bibinfo{year}{2018}).

\bibitem{barnardDeliveryStableUltrathin2022}
\bibinfo{author}{Barnard, J. C.~T.} \emph{et~al.}
\newblock \bibinfo{title}{Delivery of stable ultra-thin liquid sheets in vacuum
  for biochemical spectroscopy}.
\newblock \emph{\bibinfo{journal}{Frontiers in Molecular Biosciences}}
  \textbf{\bibinfo{volume}{9}} (\bibinfo{year}{2022}).

\bibitem{ferchaudInteractionIntenseFewcycle2022}
\bibinfo{author}{Ferchaud, C.} \emph{et~al.}
\newblock \bibinfo{title}{Interaction of an intense few-cycle infrared laser
  pulse with an ultrathin transparent liquid sheet}.
\newblock \emph{\bibinfo{journal}{Optics Express}}
  \textbf{\bibinfo{volume}{30}}, \bibinfo{pages}{34684--34692}
  (\bibinfo{year}{2022}).

\bibitem{ayusoSyntheticChiralLight2019}
\bibinfo{author}{Ayuso, D.} \emph{et~al.}
\newblock \bibinfo{title}{Synthetic chiral light for efficient control of
  chiral light--matter interaction}.
\newblock \emph{\bibinfo{journal}{Nature Photonics}}
  \textbf{\bibinfo{volume}{13}}, \bibinfo{pages}{866} (\bibinfo{year}{2019}).

\bibitem{brixnerQuantumControlGasPhase2003}
\bibinfo{author}{Brixner, T.} \& \bibinfo{author}{Gerber, G.}
\newblock \bibinfo{title}{Quantum {{Control}} of {{Gas-Phase}} and
  {{Liquid-Phase Femtochemistry}}}.
\newblock \emph{\bibinfo{journal}{ChemPhysChem}} \textbf{\bibinfo{volume}{4}},
  \bibinfo{pages}{418--438} (\bibinfo{year}{2003}).

\bibitem{cirmiOpticalWaveformSynthesis2023}
\bibinfo{author}{Cirmi, G.} \emph{et~al.}
\newblock \bibinfo{title}{Optical {{Waveform Synthesis}} and {{Its
  Applications}}}.
\newblock \emph{\bibinfo{journal}{Laser \& Photonics Reviews}}
  \textbf{\bibinfo{volume}{17}}, \bibinfo{pages}{2200588}
  (\bibinfo{year}{2023}).

\bibitem{muckeWaveformNonlinearOptics2015}
\bibinfo{author}{M{\"u}cke, O.~D.} \emph{et~al.}
\newblock \bibinfo{title}{Toward {{Waveform Nonlinear Optics Using
  Multimillijoule Sub-Cycle Waveform Synthesizers}}}.
\newblock \emph{\bibinfo{journal}{IEEE Journal of Selected Topics in Quantum
  Electronics}} \textbf{\bibinfo{volume}{21}}, \bibinfo{pages}{1--12}
  (\bibinfo{year}{2015}).

\bibitem{alismailMultioctaveCEPstableSource2020}
\bibinfo{author}{Alismail, A.} \emph{et~al.}
\newblock \bibinfo{title}{Multi-octave, {{CEP-stable}} source for high-energy
  field synthesis}.
\newblock \emph{\bibinfo{journal}{Science Advances}}
  \textbf{\bibinfo{volume}{6}}, \bibinfo{pages}{eaax3408}
  (\bibinfo{year}{2020}).

\bibitem{xueCustomTailoredMultiTWOptical2021}
\bibinfo{author}{Xue, B.} \emph{et~al.}
\newblock \bibinfo{title}{A {{Custom-Tailored Multi-TW Optical Electric Field}}
  for {{Gigawatt Soft-X-Ray Isolated Attosecond Pulses}}}.
\newblock \emph{\bibinfo{journal}{Ultrafast Science}}
  \textbf{\bibinfo{volume}{2021}} (\bibinfo{year}{2021}).

\bibitem{alqattanAttosecondLightField2022}
\bibinfo{author}{Alqattan, H.}, \bibinfo{author}{Hui, D.},
  \bibinfo{author}{Pervak, V.} \& \bibinfo{author}{Hassan, M.~T.}
\newblock \bibinfo{title}{Attosecond light field synthesis}.
\newblock \emph{\bibinfo{journal}{APL Photonics}} \textbf{\bibinfo{volume}{7}},
  \bibinfo{pages}{041301} (\bibinfo{year}{2022}).

\bibitem{ridenteElectroopticCharacterizationSynthesized2022}
\bibinfo{author}{Ridente, E.} \emph{et~al.}
\newblock \bibinfo{title}{Electro-optic characterization of synthesized
  infrared-visible light fields}.
\newblock \emph{\bibinfo{journal}{Nature Communications}}
  \textbf{\bibinfo{volume}{13}}, \bibinfo{pages}{1111} (\bibinfo{year}{2022}).

\bibitem{lambertOpticalActivityCarvone2012}
\bibinfo{author}{Lambert, J.}, \bibinfo{author}{Compton, R.~N.} \&
  \bibinfo{author}{Crawford, T.~D.}
\newblock \bibinfo{title}{The optical activity of carvone: {{A}} theoretical
  and experimental investigation}.
\newblock \emph{\bibinfo{journal}{The Journal of Chemical Physics}}
  \textbf{\bibinfo{volume}{136}}, \bibinfo{pages}{114512}
  (\bibinfo{year}{2012}).

\bibitem{balavoinePreparationChiralCompounds1974}
\bibinfo{author}{Balavoine, G.}, \bibinfo{author}{Moradpour, A.} \&
  \bibinfo{author}{Kagan, H.~B.}
\newblock \bibinfo{title}{Preparation of chiral compounds with high optical
  purity by irradiation with circularly polarized light, a model reaction for
  the prebiotic generation of optical activity}.
\newblock \emph{\bibinfo{journal}{Journal of the American Chemical Society}}
  \textbf{\bibinfo{volume}{96}}, \bibinfo{pages}{5152--5158}
  (\bibinfo{year}{1974}).

\bibitem{meinertPhotonenergyControlledSymmetryBreaking2014}
\bibinfo{author}{Meinert, C.} \emph{et~al.}
\newblock \bibinfo{title}{Photonenergy-{{Controlled Symmetry Breaking}} with
  {{Circularly Polarized Light}}}.
\newblock \emph{\bibinfo{journal}{Angewandte Chemie}}
  \textbf{\bibinfo{volume}{126}}, \bibinfo{pages}{214--218}
  (\bibinfo{year}{2014}).

\bibitem{bauerCatalyticEnantioselectiveReactions2005}
\bibinfo{author}{Bauer, A.}, \bibinfo{author}{Westk{\"a}mper, F.},
  \bibinfo{author}{Grimme, S.} \& \bibinfo{author}{Bach, T.}
\newblock \bibinfo{title}{Catalytic enantioselective reactions driven by
  photoinduced electron transfer}.
\newblock \emph{\bibinfo{journal}{Nature}} \textbf{\bibinfo{volume}{436}},
  \bibinfo{pages}{1139--1140} (\bibinfo{year}{2005}).

\bibitem{wanieCapturingElectrondrivenChiral2024a}
\bibinfo{author}{Wanie, V.} \emph{et~al.}
\newblock \bibinfo{title}{Capturing electron-driven chiral dynamics in
  {{UV-excited}} molecules}.
\newblock \emph{\bibinfo{journal}{Nature}} \textbf{\bibinfo{volume}{630}},
  \bibinfo{pages}{109} (\bibinfo{year}{2024}).

\bibitem{huckDynamicControlAmplification1996}
\bibinfo{author}{Huck, N. P.~M.}, \bibinfo{author}{Jager, W.~F.},
  \bibinfo{author}{de~Lange, B.} \& \bibinfo{author}{Feringa, B.~L.}
\newblock \bibinfo{title}{Dynamic {{Control}} and {{Amplification}} of
  {{Molecular Chirality}} by {{Circular Polarized Light}}}.
\newblock \emph{\bibinfo{journal}{Science}} \textbf{\bibinfo{volume}{273}},
  \bibinfo{pages}{1686--1688} (\bibinfo{year}{1996}).

\bibitem{kohnDensityFunctionalTheory1996}
\bibinfo{author}{Kohn, W.}, \bibinfo{author}{Becke, A.~D.} \&
  \bibinfo{author}{Parr, R.~G.}
\newblock \bibinfo{title}{Density {{Functional Theory}} of {{Electronic
  Structure}}}.
\newblock \emph{\bibinfo{journal}{The Journal of Physical Chemistry}}
  \textbf{\bibinfo{volume}{100}}, \bibinfo{pages}{12974--12980}
  (\bibinfo{year}{1996}).

\bibitem{zieglerApproximateDensityFunctional2002}
\bibinfo{author}{Ziegler, T.}
\newblock \bibinfo{title}{Approximate density functional theory as a practical
  tool in molecular energetics and dynamics}.
\newblock \emph{\bibinfo{journal}{Chemical Reviews}}
  \textbf{\bibinfo{volume}{91}}, \bibinfo{pages}{651--667}
  (\bibinfo{year}{1991}).

\bibitem{leeDevelopmentColleSalvettiCorrelationenergy1988}
\bibinfo{author}{Lee, C.}, \bibinfo{author}{Yang, W.} \& \bibinfo{author}{Parr,
  R.~G.}
\newblock \bibinfo{title}{Development of the {{Colle-Salvetti}}
  correlation-energy formula into a functional of the electron density}.
\newblock \emph{\bibinfo{journal}{Physical Review B}}
  \textbf{\bibinfo{volume}{37}}, \bibinfo{pages}{785--789}
  (\bibinfo{year}{1988}).

\bibitem{beckeDensityfunctionalExchangeenergyApproximation1988}
\bibinfo{author}{Becke, A.~D.}
\newblock \bibinfo{title}{Density-functional exchange-energy approximation with
  correct asymptotic behavior}.
\newblock \emph{\bibinfo{journal}{Physical Review A}}
  \textbf{\bibinfo{volume}{38}}, \bibinfo{pages}{3098--3100}
  (\bibinfo{year}{1988}).

\bibitem{mcleanContractedGaussianBasis2008}
\bibinfo{author}{McLean, A.~D.} \& \bibinfo{author}{Chandler, G.~S.}
\newblock \bibinfo{title}{Contracted {{Gaussian}} basis sets for molecular
  calculations. {{I}}. {{Second}} row atoms, {{Z}}=11--18}.
\newblock \emph{\bibinfo{journal}{The Journal of Chemical Physics}}
  \textbf{\bibinfo{volume}{72}}, \bibinfo{pages}{5639--5648}
  (\bibinfo{year}{2008}).

\bibitem{g16}
\bibinfo{author}{Frisch, M.~J.} \emph{et~al.}
\newblock \bibinfo{title}{Gaussian 16 {R}evision {C}.01}
  (\bibinfo{year}{2016}).
\newblock \bibinfo{note}{Gaussian Inc. Wallingford CT}.

\bibitem{luMultiwfnMultifunctionalWavefunction2012}
\bibinfo{author}{Lu, T.} \& \bibinfo{author}{Chen, F.}
\newblock \bibinfo{title}{Multiwfn: {{A}} multifunctional wavefunction
  analyzer}.
\newblock \emph{\bibinfo{journal}{Journal of Computational Chemistry}}
  \textbf{\bibinfo{volume}{33}}, \bibinfo{pages}{580--592}
  (\bibinfo{year}{2012}).

\bibitem{toroTwoPhotonAbsorptionCircular2010}
\bibinfo{author}{Toro, C.} \emph{et~al.}
\newblock \bibinfo{title}{Two-{{Photon Absorption Circular Dichroism}}: {{A New
  Twist}} in {{Nonlinear Spectroscopy}}}.
\newblock \emph{\bibinfo{journal}{Chemistry -- A European Journal}}
  \textbf{\bibinfo{volume}{16}}, \bibinfo{pages}{3504--3509}
  (\bibinfo{year}{2010}).

\bibitem{toroTwophotonAbsorptionCircularlinear2010}
\bibinfo{author}{Toro, C.} \emph{et~al.}
\newblock \bibinfo{title}{Two-photon absorption circular-linear dichroism on
  axial enantiomers}.
\newblock \emph{\bibinfo{journal}{Chirality}} \textbf{\bibinfo{volume}{22}},
  \bibinfo{pages}{E202--E210} (\bibinfo{year}{2010}).

\bibitem{linVibronicallyResolvedElectronic2008}
\bibinfo{author}{Lin, N.}, \bibinfo{author}{Santoro, F.},
  \bibinfo{author}{Zhao, X.}, \bibinfo{author}{Rizzo, A.} \&
  \bibinfo{author}{Barone, V.}
\newblock \bibinfo{title}{Vibronically {{Resolved Electronic Circular Dichroism
  Spectra}} of ({{R}})-(+)-3-{{Methylcyclopentanone}}: {{A Theoretical
  Study}}}.
\newblock \emph{\bibinfo{journal}{The Journal of Physical Chemistry A}}
  \textbf{\bibinfo{volume}{112}}, \bibinfo{pages}{12401--12411}
  (\bibinfo{year}{2008}).

\bibitem{linTheoryVibrationallyResolved2009}
\bibinfo{author}{Lin, N.} \emph{et~al.}
\newblock \bibinfo{title}{Theory for {{Vibrationally Resolved Two-Photon
  Circular Dichroism Spectra}}. {{Application}} to
  ({{R}})-(+)-3-{{Methylcyclopentanone}}}.
\newblock \emph{\bibinfo{journal}{The Journal of Physical Chemistry A}}
  \textbf{\bibinfo{volume}{113}}, \bibinfo{pages}{4198--4207}
  (\bibinfo{year}{2009}).

\bibitem{rizzoInitioStudyExcited2011}
\bibinfo{author}{Rizzo, A.} \& \bibinfo{author}{Vahtras, O.}
\newblock \bibinfo{title}{Ab initio study of excited state electronic circular
  dichroism. {{Two}} prototype cases: {{Methyl}} oxirane and
  {{R-}}(+)-1,1{$\prime$}-bi(2-naphthol)}.
\newblock \emph{\bibinfo{journal}{The Journal of Chemical Physics}}
  \textbf{\bibinfo{volume}{134}}, \bibinfo{pages}{244109}
  (\bibinfo{year}{2011}).

\bibitem{andersenProbingMolecularChirality2022}
\bibinfo{author}{Andersen, J.~H.}, \bibinfo{author}{Nanda, K.~D.},
  \bibinfo{author}{Krylov, A.~I.} \& \bibinfo{author}{Coriani, S.}
\newblock \bibinfo{title}{Probing {{Molecular Chirality}} of {{Ground}} and
  {{Electronically Excited States}} in the {{UV}}--vis and {{X-ray Regimes}}:
  {{An EOM-CCSD Study}}}.
\newblock \emph{\bibinfo{journal}{Journal of Chemical Theory and Computation}}
  \textbf{\bibinfo{volume}{18}}, \bibinfo{pages}{1748--1764}
  (\bibinfo{year}{2022}).

\bibitem{ordonezGeometricApproachDecoding2022}
\bibinfo{author}{Ordonez, A.~F.} \& \bibinfo{author}{Smirnova, O.}
\newblock \bibinfo{title}{A geometric approach to decoding molecular structure
  and dynamics from photoionization of isotropic samples}.
\newblock \emph{\bibinfo{journal}{Physical Chemistry Chemical Physics}}
  \textbf{\bibinfo{volume}{24}}, \bibinfo{pages}{13605} (\bibinfo{year}{2022}).

\bibitem{andrewsThreeDimensionalRotational1977}
\bibinfo{author}{Andrews, D.~L.} \& \bibinfo{author}{Thirunamachandran, T.}
\newblock \bibinfo{title}{On three-dimensional rotational averages}.
\newblock \emph{\bibinfo{journal}{The Journal of Chemical Physics}}
  \textbf{\bibinfo{volume}{67}}, \bibinfo{pages}{5026--5033}
  (\bibinfo{year}{1977}).

\end{thebibliography}

\end{document}